\documentclass{article}
\usepackage{spconf,amsmath,epsfig}
\usepackage{color}
\usepackage{bm}

\usepackage{multicol}
\usepackage{multirow}
\usepackage{balance}
\usepackage{hyperref}
\usepackage{graphicx} 
\usepackage{epstopdf}
\usepackage{enumitem}
\usepackage{hyperref}
\setenumerate[1]{itemsep=0pt,partopsep=0pt,parsep=\parskip,topsep=5pt}
\setitemize[1]{itemsep=0pt,partopsep=0pt,parsep=\parskip,topsep=5pt}
\setdescription{itemsep=0pt,partopsep=0pt,parsep=\parskip,topsep=5pt}
\begin{document}\sloppy

\def\etal{{\emph{et al}}}

\title{Physical Model Guided Deep Image Deraining}
%
\name{Honghe Zhu$^{\triangle, 1}$\thanks{$\triangle$ denotes that Honghe Zhu and Cong Wang have equal contributions to this work.}, Cong Wang$^{\triangle, 1}$, Yajie Zhang$^{2}$, Zhixun Su$^{*, 1, 3}$\thanks{* denotes the corresponding author: zxsu@dlut.edu.cn.}, Guohui Zhao$^{1}$}
\address{$^{1}$ Dalian University of Technology\\
$^{2}$ University of Chinese Academy of Sciences
\\
$^{3}$ Guilin University of Electronic Technology
}

\maketitle

\begin{abstract}
  Single image deraining is an urgent task because the degraded rainy image makes many computer vision systems fail to work, such as video surveillance and autonomous driving.
  So, deraining becomes important and an effective deraining algorithm is needed.
  In this paper, we propose a novel network based on physical model guided learning for single image deraining, which consists of three sub-networks: rain streaks network, rain-free network, and guide-learning network.
  The concatenation of rain streaks and rain-free image that are estimated by rain streaks network, rain-free network, respectively, is input to the guide-learning network to guide further learning and the direct sum of the two estimated images is constrained with the input rainy image based on the physical model of rainy image.
  Moreover, we further develop the Multi-Scale Residual Block (MSRB) to better utilize multi-scale information and it is proved to boost the deraining performance.
  Quantitative and qualitative experimental results demonstrate that the proposed method outperforms the state-of-the-art deraining methods.
  The source code will be available at \url{https://supercong94.wixsite.com/supercong94}.
\end{abstract}
\vspace{-2mm}
\section{Introduction}
\label{sec: Introduction}
Rain is a very common weather phenomenon, and images and videos captured in rain consist of raindrops and rain streaks with different speeds, different directions and various density levels, which causes many computer vision systems likely fail to work.
So, removing the rain components from rainy images or videos, which obtains a clear background scene, is needed.
There are two categories of deraining: single image-based methods~\cite{derain_dsc_luo,derain_id_kang,derain_lowrank,derain_lp_li,derain_clearing_fu,derain_ddn_fu,derain_jorder_yang} and video-based methods~\cite{derain_prior_Garg,videoderain,derain_prior,videoderain_Tripathi}.
As the temporal information can be leveraged by analyzing the difference between adjacent frames in a video, hence, video-based methods are easier than single image-based methods.
In this paper, we explore the more difficult problem, single image deraining.

Image deraining has attracted much attention in recent years, which is always based on this physical rainy model: the observed rainy image is generally modeled as a linear sum of a rain-free background image and rain streaks.
In the mathematical representation, the model can be expressed as:
\begin{equation}
\bm{O} = \bm{B} + \bm{R},
\label{eq:rain model}
\end{equation}
where $\bm O$, $\bm B$, and $\bm R$ denote the observed rainy images, clear background images, and rain streaks, respectively.
Based on the Eq.~\eqref{eq:rain model}, deraining methods should remove $\bm R$ from $\bm O$ to get $\bm B$, which is a highly ill-posed problem, due to there are a series of solutions of $\bm B$, $\bm R$ for a given $\bm O$, theoretically.

To make the problem well be solved, numerous conventional methods adopt various priors about rain streaks or clean background scene to constrain the solution space, such as the sparse code~\cite{derain_dsc_luo}, image decomposition~\cite{derain_id_kang}, low-rank~\cite{derain_lowrank} and Gaussian mixture model~\cite{derain_lp_li}.
These deraining methods always make simple hypotheses on $\bm R$, i.e. rain streaks, such as the assumptions that the rain streaks are sparse and have similar characters in falling directions and shapes, which only work in some specific cases.

With the rise of deep learning, numerous methods have achieved greatly succeeded in many computer vision tasks~\cite{pyramid_Pose_Estimation,semanticsegmentation_fcn,Pyramid_Object_Detection} due to the powerful feature representation capability.
Deraining methods also acquire significantly improvement via these deep learning-based methods~\cite{derain_huang_context,derain_cgan_zhang,derain_clearing_fu,derain_ddn_fu,derain_rescan_li}.
However, they still exist some limitations.

On the one hand, many existing methods usually only estimate the rain streak or rain-free image~\cite{derain_rescan_li,derain_ddn_fu,derain_jorder_yang}, and they neglect that the estimated rain streaks and rain-free image can serve as a physical model guide for the deraining process.
On the other hand, multi-scale operations can better acquire the rain streaks information with different levels, which should have a boost effect for deraining.
However, numerous deep learning-based methods~\cite{derain_ddn_fu,derain_cgan_zhang,derain_rescan_li} do not consider the effect of multi-scale information into deraining.

To handle with above limitations, we propose a novel network based on physical model guided learning that utilizes physical model to guide the learning process and applies the multi-scale manner into feature maps.
Specifically, the sum of the estimated rain streaks and rain-free image is compared with their corresponding rainy image as a constraint term according to the rainy physical model~\ref{eq:rain model} and the concatenation of them is input into guide-learning as a guide to learn.
Moreover, we design a Multi-Scale Residual Block (MSRB) to obtain different features with different levels.

Our contributions are summarized as followings:
\begin{itemize}
\item We design the guide-learning network based on the rainy physical model and the guide boost the deraining performance on both details and texture information.
\item We propose a Multi-Scale Residual Block (MSRB) to better utilize multi-scale information and experiments prove that the block is favorable for improving the rain streaks representation capability.
\item Our proposed network outperforms the state-of-the-art methods on synthetic and real-world datasets in visually, quantitatively and qualitatively.
\end{itemize}

%
\vspace{-2mm}
\section{Related Work}

In this section, we present a brief review on single image deraining approaches that can be split into prior-based methods and deep learning-based methods.

For prior based methods, Kang \etal.~\cite{derain_id_kang} first decomposed the rainy image into a low- and high-frequency layer, and then utilized sparse coding to remove the rain streaks in high-frequency layer.
Chen \etal.~\cite{derain_lowrank} assumed the rain steaks are low-rank and proposed an effective low-rank structure to model rain streaks.
Luo \etal.~\cite{derain_dsc_luo} proposed a discriminative sparse coding framework to accurately separate rain streaks and clean background scenes.
Li \etal.~\cite{derain_lp_li} used patch priors based on Gaussian Mixture Models for both rain steaks and clear background to remove the rain streaks.

For deep learning-based methods, Fu \etal.~\cite{derain_clearing_fu,derain_ddn_fu} first applied deep learning in single image deraining that they decompose rainy image into low- and high-frequency parts, and then put the high-frequency part into a Convolutional Neural Network (CNN) to estimate residual rain streaks.
Yang \etal.~\cite{derain_jorder_yang} proposed a recurrent contextual network that can jointly detect and remove rain steaks.
Zhang \etal.~\cite{derain_cgan_zhang} designed a generative adversarial network to prevent the degeneration of background image and utilized perceptual loss to refine visual quality.
Fan \etal.~\cite{derain_GRN} generalized a residual-guide network for deraining.
Li \etal.~\cite{derain_rescan_li} utilized squeeze-and-excitation to learn different weights of different rain streaks layer for deraining.
Ren~\etal.~\cite{derain_prenet_Ren_2019_CVPR} considered network architecture, input and output, and loss functions and provided a better and simpler baseline deraining network.
\vspace{-2mm}
\section{Proposed Method}
\label{sec: Proposed Method}
%
\begin{figure}[!t]
\begin{center}
\begin{tabular}{c}
\includegraphics[width = 0.98\linewidth]{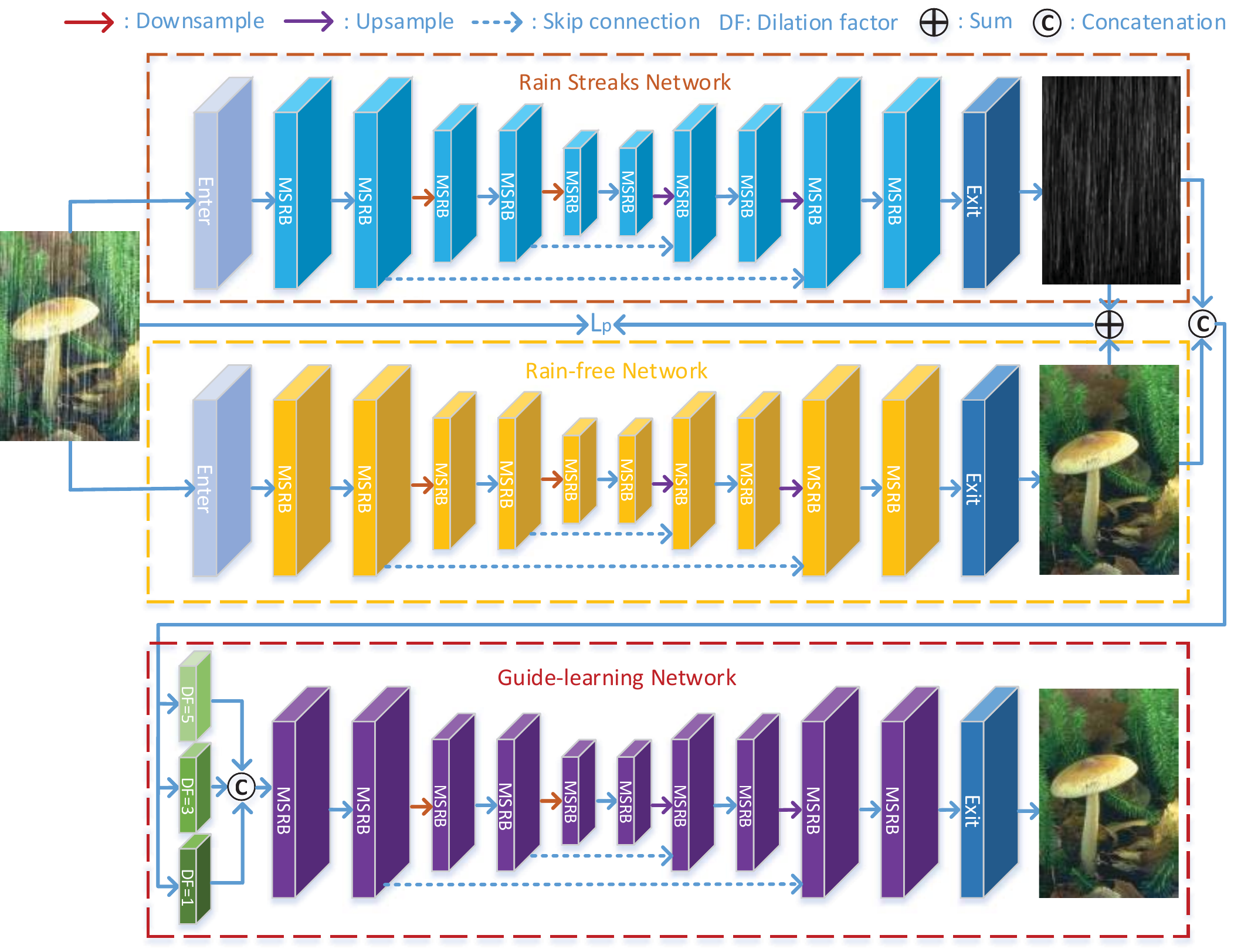}
\end{tabular}
\end{center}
\vspace{-6mm}
\caption{Overall Network Framework. MSRB is shown in Fig~\ref{fig: MSRB}.
The overall network consists of three sub-networks: Rain Streaks Network, Rain-free Network, and Guide-learning Network.
The Rain Streak Network and Rain-free Network learn to estimate rain streaks and rain-free images, respectively, and their outputs are cascaded to input the Guide-learning Network as the further guided learning.
}
\vspace{-5mm}
\label{fig: Overall Framework}
\end{figure}
\label{sec: Related Work}
In this section, we state more details about our proposed method, including its overall network framework, the multi-scale residual block (MSRB) and loss functions.

\subsection{Overall framework}
\label{sec: Overall framework}
As shown in Fig.~\ref{fig: Overall Framework}, the proposed network consists of three sub-networks: rain streaks network, rain-free network, and guide-learning network.
The first two sub-networks have the same structures that are both encoder-decoder.
And in order to learn better spatial contextual information to further guide to restore clear image, the estimated rain streaks and rain-free image are cascaded to input the guide-learning network with multi-stream dilation convolution to further refine the deraining results.
Moreover, to restrain the rain streaks network and rain-free network to generate better according results, the add between estimated rain streaks and rain-free images is restrained via $L_{1}$ norm according to rainy physical model~\ref{eq:rain model}.
Furthermore, MSRB is designed to acquire multi-scale information by combining the multi-scale operations and residual block.

\subsection{Multi-Scale Residual Block (MSRB)}
\label{sec: Multi-Scale Residual Block}
\begin{figure}[!h]
\begin{center}
\begin{tabular}{c}
\includegraphics[width = 0.6\linewidth]{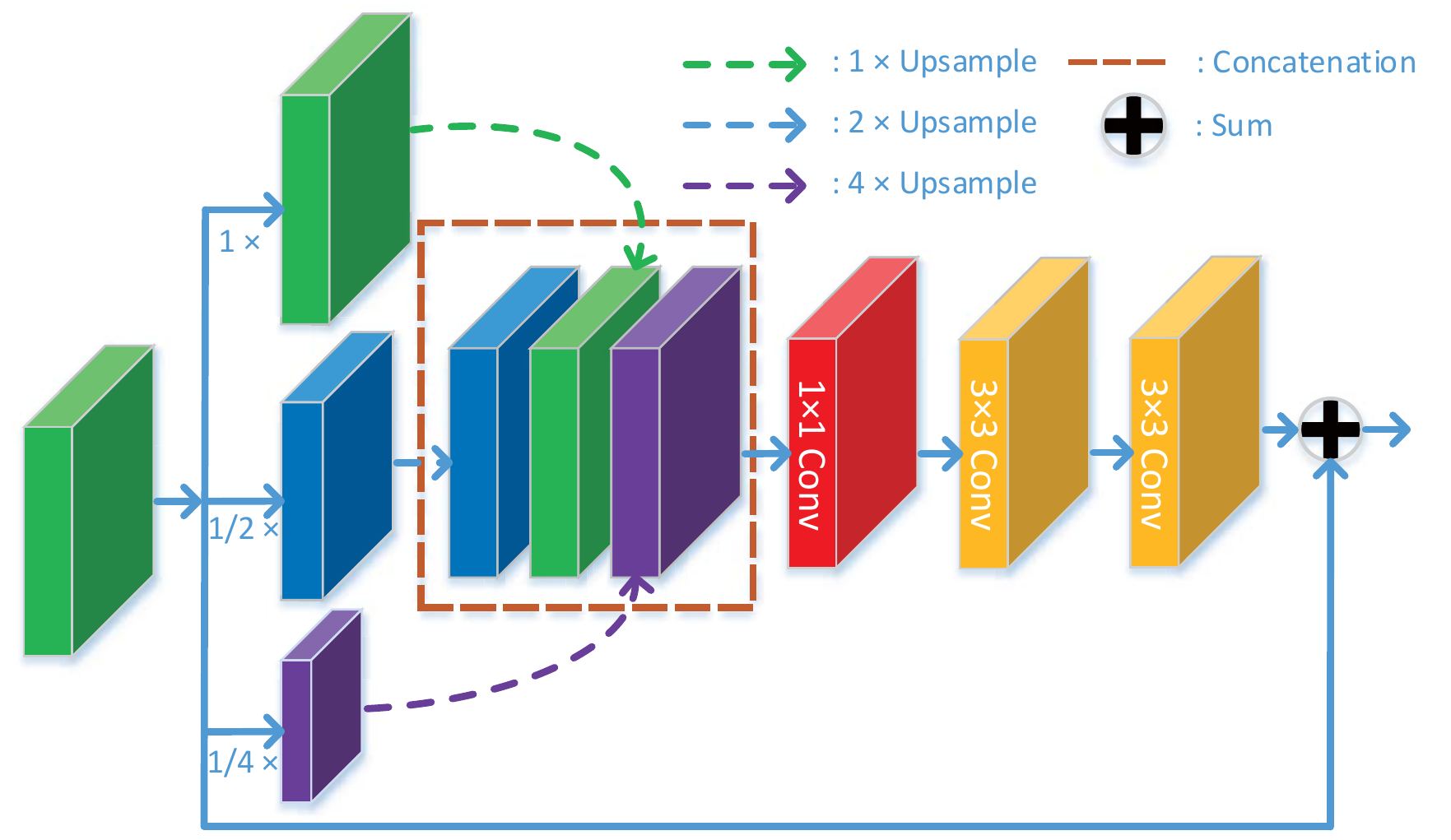}
\end{tabular}
\end{center}
\vspace{-6mm}
\caption{Multi-Scale Residual Block (MSRB).}
\label{fig: MSRB}
\end{figure}

Multi-scale features have been widely leveraged in many computer vision systems, such as face-alignment~\cite{encoder-decoder}, semantic segmentation~\cite{multi_scale_scene_parsing}, depth estimation~\cite{Depth_map_prediction} and single image super-resolution~\cite{Residual_dense_resolution}.
Combining features at different scales can result in a better representation of an object and its surrounding context.
Therefore, multi-scale residual block (MSRB) is proposed that is the concatenation between different scales of feature maps and the residual block, as shown in Fig.~\ref{fig: MSRB}.

We describe the MSRB mathematically:
Firstly, we utilize $Pooling$ operation with different size of kernels and strides to obtain the multi-scale features:
\begin{equation}
y_{i} = Pooling_{i}(x), i = 1, 2, 4.
\label{eq:y}
\end{equation}
where $Pooling_{i}$ denotes $Pooling$ operation with $i \times i$ kernel and stride.
\\
Lastly, all the scales are fused and feed into three convolution layers then add the original input signal $x$ to learn the residual:
\begin{equation}
z = H(Cat[Up_{1}(y_{1}), \cdots, Up_{4}(y_{4})]) + x.
\label{eq:yr}
\end{equation}
where $Up_{i}$ denotes $i \times$ Upsampling operation and $Cat$ denotes concatenation operation at the channel dimension.
$H$ denotes a series of operations that consist of two $3 \times 3$ and one $1 \times 1$ convolution operations.
The MSRB can learn features with different scales and all different features are fused to learn the primary feature.

\subsection{Loss function}
\label{sec: Loss function}
We use $L_1$-norm as the loss function.

For the rain streaks network and rain-free network:
\begin{equation}
L_{rain} = \lVert \widetilde{\bm R} - {\bm R} \rVert_{1},
\label{eq:Rain}
\end{equation}
\begin{equation}
L_{rain-free} = \lVert \widetilde{\bm B} - {\bm B} \rVert_{1},
\label{eq:image}
\end{equation}
where $\widetilde{\bm R}$, $\widetilde{\bm B}$ denote the estimated rain streaks layer and clean background image, ${\bm R}$ and ${\bm B}$ denote the ground truth of rain streaks and rain-free image.
For guide-learning network:
\begin{equation}
L_{guide} = \lVert \widehat{\bm B} - {\bm B} \rVert_{1},
\label{eq:guide}
\end{equation}
where $\widehat{\bm B}$ denote the output of guide-learning network, i.e. the final estimated rain-free image.

Moreover, we compute the $L_1$-norm of the input rainy image $\bm{O}$ and the sum of $\widetilde{\bm R}$, $\widetilde{\bm B}$ in order to constrain the solution space of rain streaks and rain-free network according to the rainy physical model~\ref{eq:rain model}:
\begin{equation}
L_{p} = \lVert \widetilde{\bm B} + \widetilde{\bm R} - \bm{O} \rVert_{1},
\label{eq:Lp}
\end{equation}

So the overall loss function is defined as:
\begin{equation}
L = L_{guide} + \alpha L_{rain} + \beta L_{rain-free} + \gamma L_{p},
\label{eq: overall loss function}
\end{equation}
where $\alpha, \beta, \gamma$ are constant.
\begin{table*}[!t]
\begin{center}
\scriptsize
\caption{Quantitative experiments evaluated on three synthetic datasets. The best results are marked in bold.}
\vspace{1mm}
\scalebox{1.4}{
\begin{tabular}{ccccccccc}
\hline
Dataset                   & Metric   & DSC~\cite{derain_dsc_luo}    & LP~\cite{derain_lp_li}     & DDN~\cite{derain_ddn_fu}        & JORDER~\cite{derain_jorder_yang}   & RESCAN~\cite{derain_rescan_li} & PReNet~\cite{derain_prenet_Ren_2019_CVPR} & \textbf{Ours}   \\ \hline
\multirow{2}{*}{Rain100H} & PSNR     & 15.66  & 14.26  & 22.26      & 23.45    & 25.92  & 27.89  & \textbf{28.96}  \\ \cline{2-9}
                          & SSIM     & 0.42   & 0.54   & 0.69       & 0.74     & 0.84   & 0.89   & \textbf{0.90} \\ \hline
\multirow{2}{*}{Rain100L} & PSNR     &24.16   & 29.11  & 34.85      & 36.11    & 36.12  & 36.69  & \textbf{38.64}  \\
\cline{2-9}
                          & SSIM     & 0.87   & 0.88   & 0.95       & 0.97     & 0.96   & 0.98   & \textbf{0.99} \\ \hline
\multirow{2}{*}{Rain1200} & PSNR     &21.44   & 22.46  & 30.95      & 29.75    & 32.35  & 32.38  & \textbf{33.42}  \\
\cline{2-9}
                          & SSIM     & 0.79   & 0.80   & 0.86       & 0.87     & 0.89   & 0.92   & \textbf{0.93} \\ \hline
\end{tabular}}
\label{tab:synthetic datasets}
\vspace{-5mm}
\end{center}
\end{table*}
%

\vspace{-2mm}
\section{Experimental Results}
\begin{figure*}[!t]
\vspace{-2mm}
\begin{center}
\begin{tabular}{ccccccc}
\includegraphics[width = 0.135\linewidth]{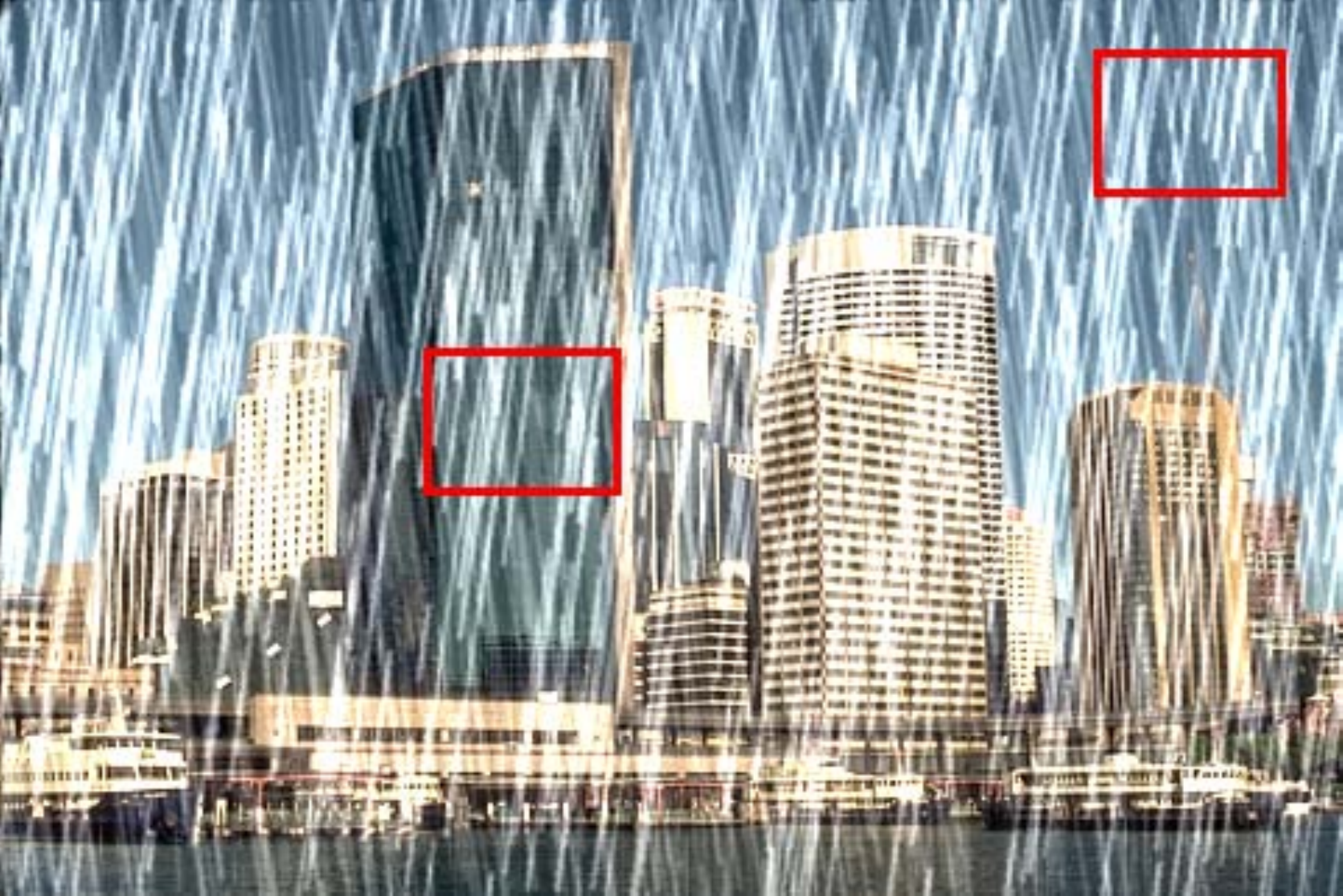} &\hspace{-4mm}
\includegraphics[width = 0.135\linewidth]{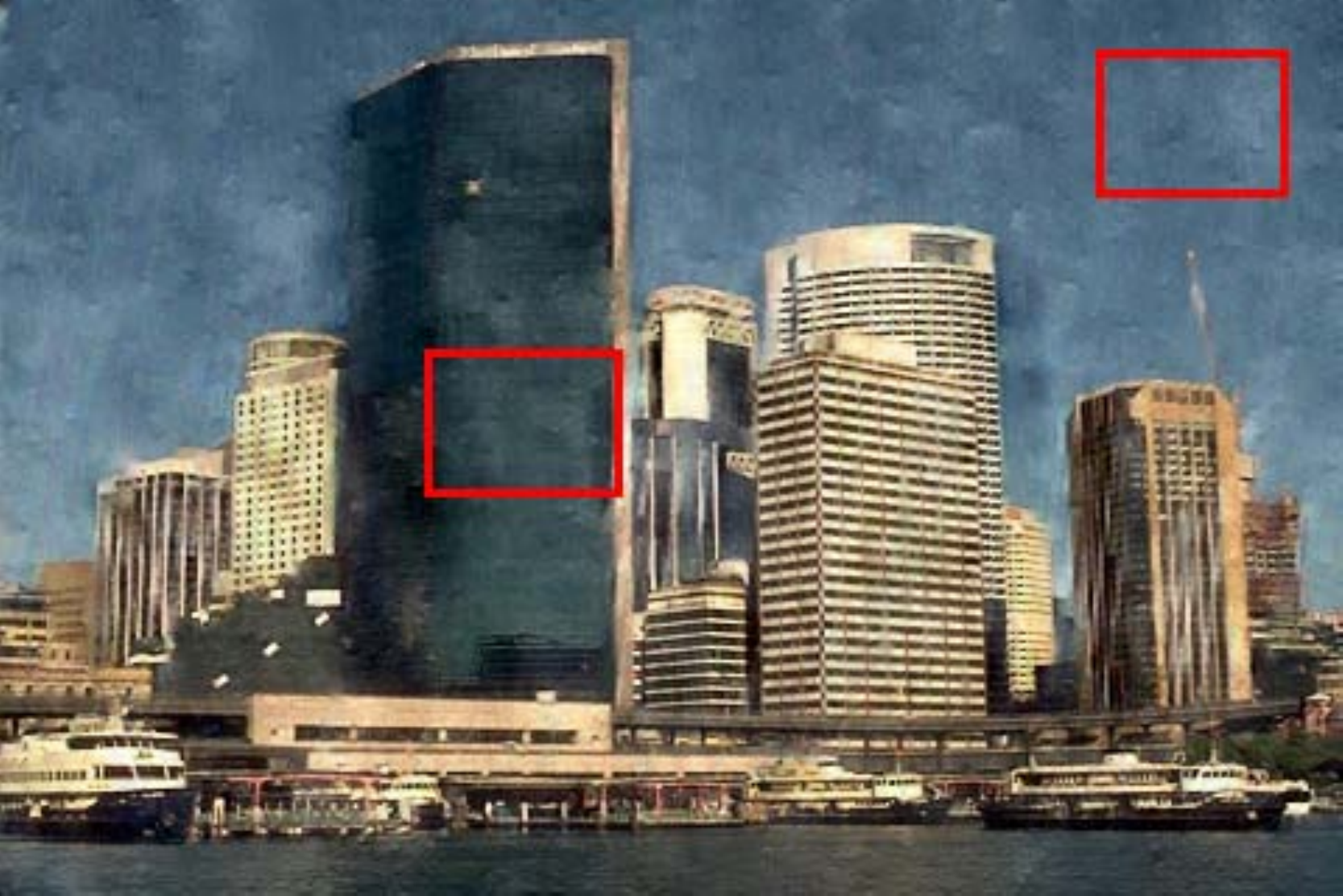} &\hspace{-4mm}
\includegraphics[width = 0.135\linewidth]{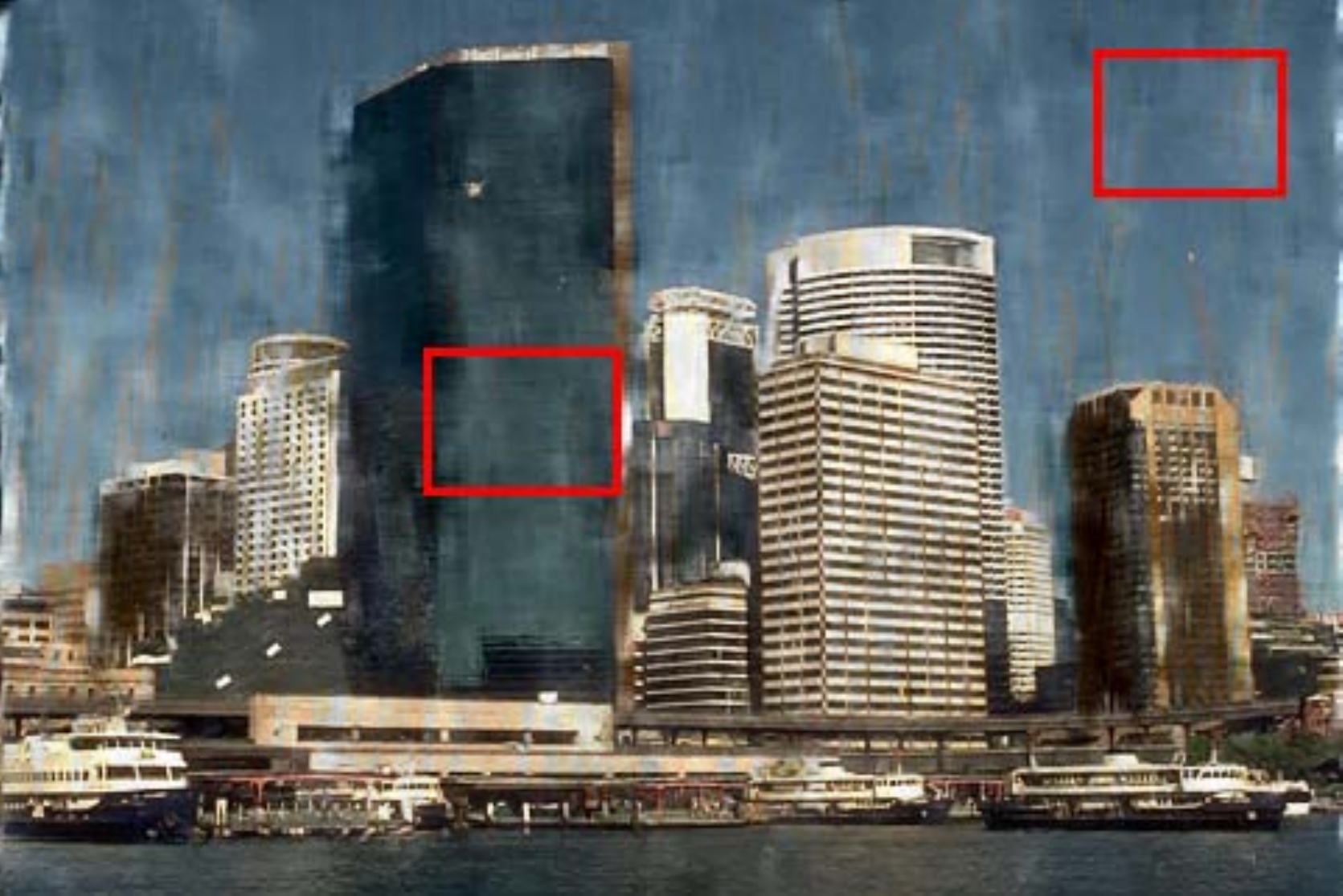} &\hspace{-4mm}
\includegraphics[width = 0.135\linewidth]{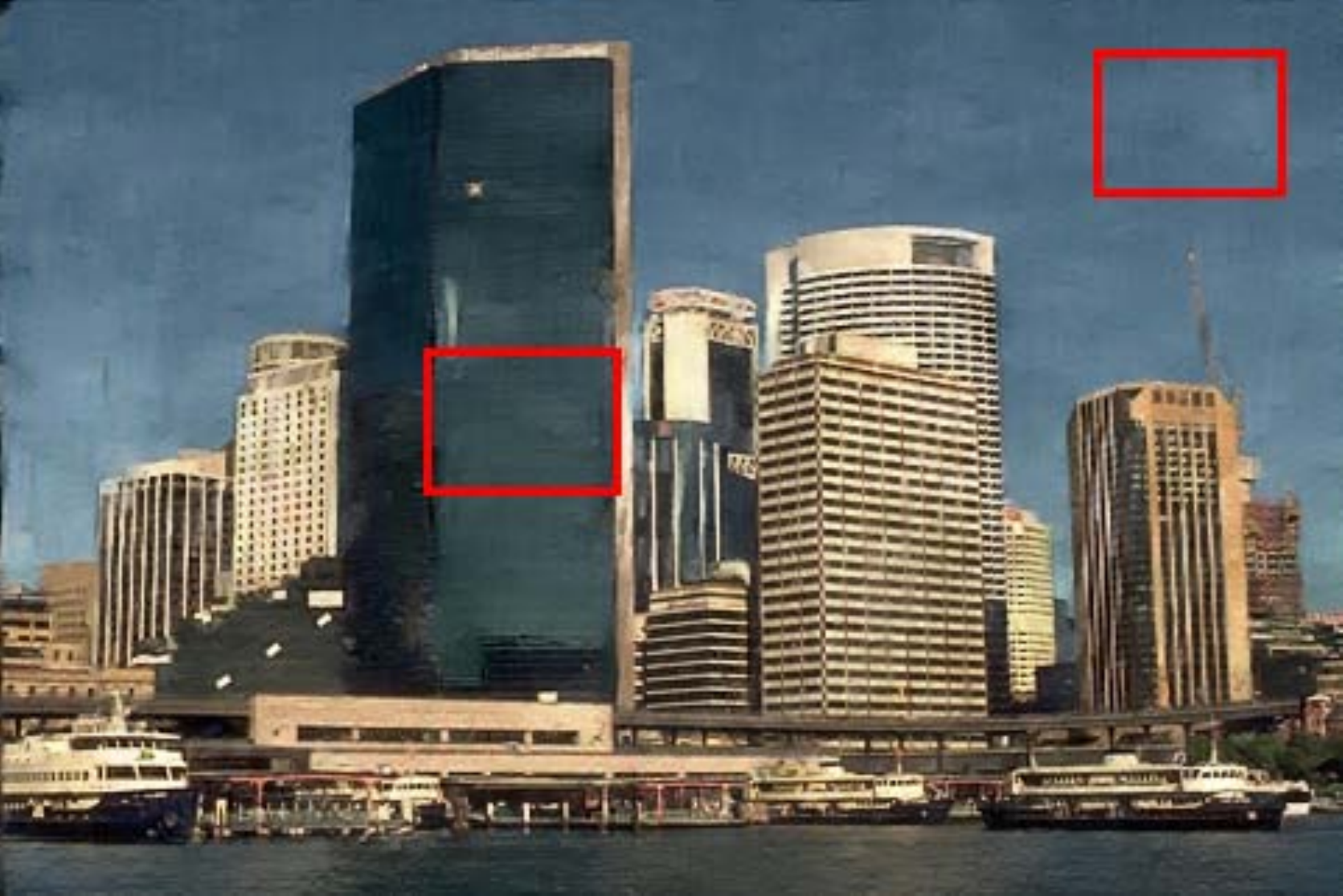} &\hspace{-4mm}
\includegraphics[width = 0.135\linewidth]{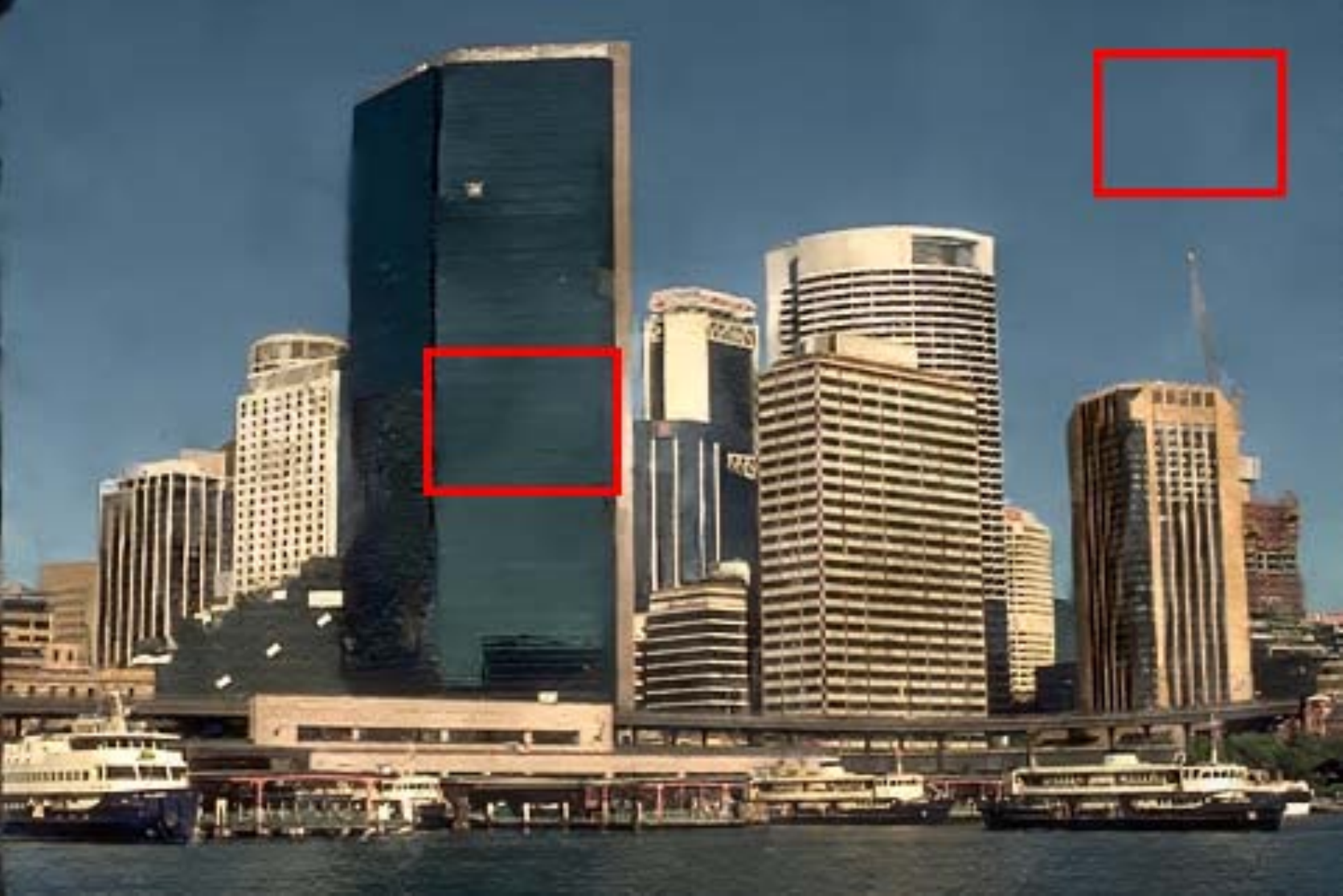} &\hspace{-4mm}
\includegraphics[width = 0.135\linewidth]{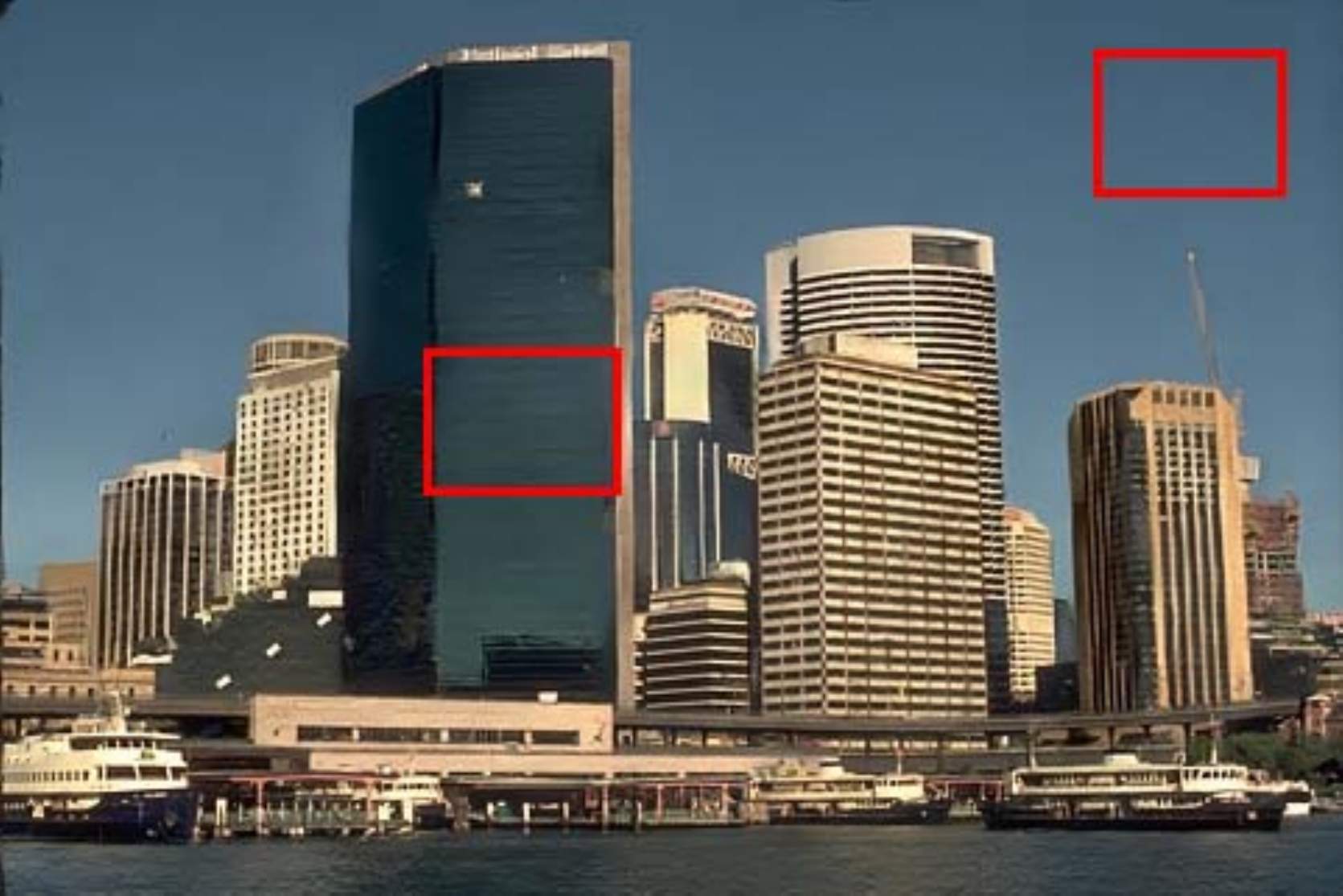} &\hspace{-4mm}
\includegraphics[width = 0.135\linewidth]{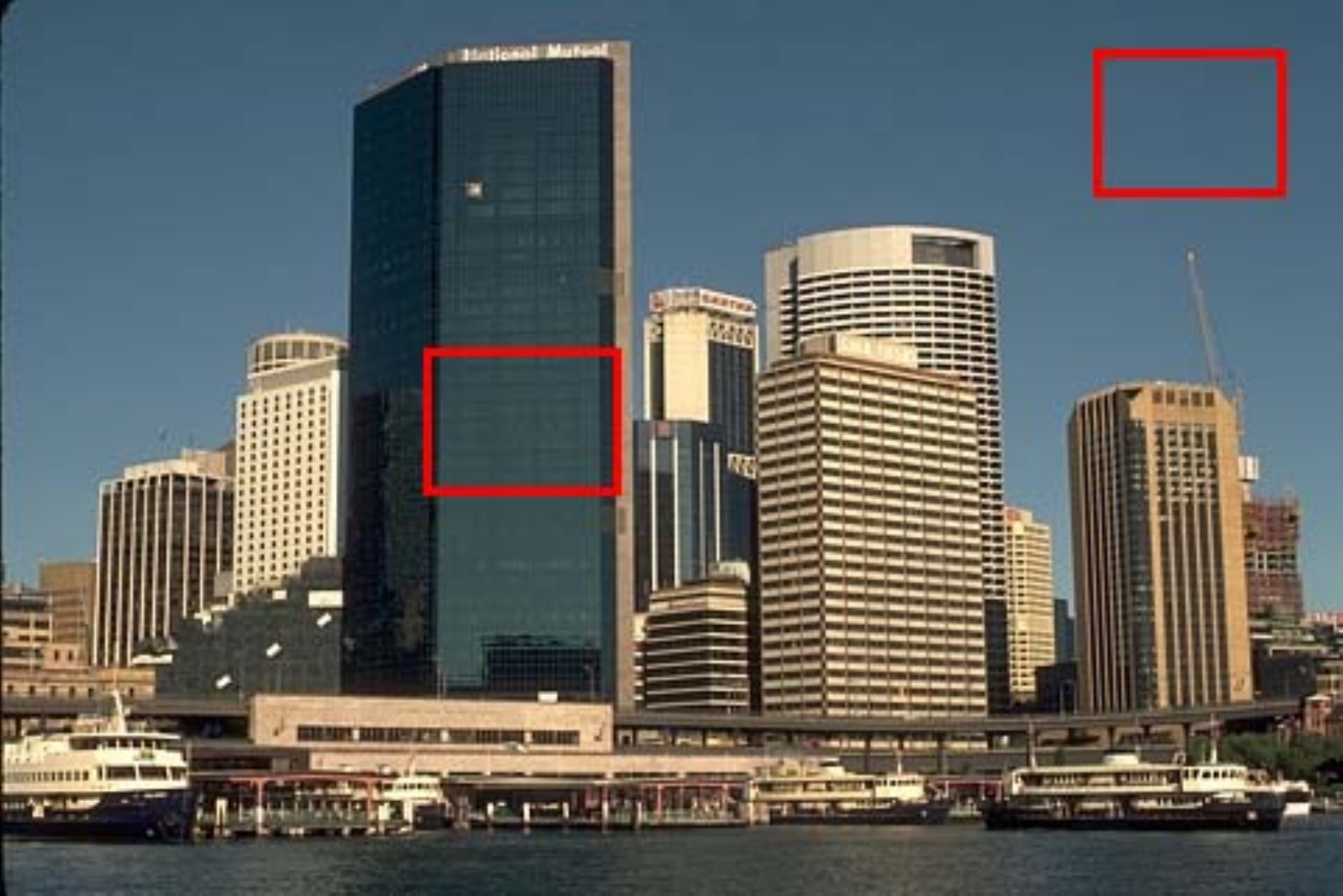}
\\
PSNR/SSIM &\hspace{-4mm}  22.12/0.79    &\hspace{-4mm}    20.31/0.75  &\hspace{-4mm} 24.49/0.79 &\hspace{-4mm}  24.93/0.92 &\hspace{-4mm}  \textbf{25.84/0.93} &\hspace{-4mm}  Inf/1
\\
\includegraphics[width = 0.135\linewidth]{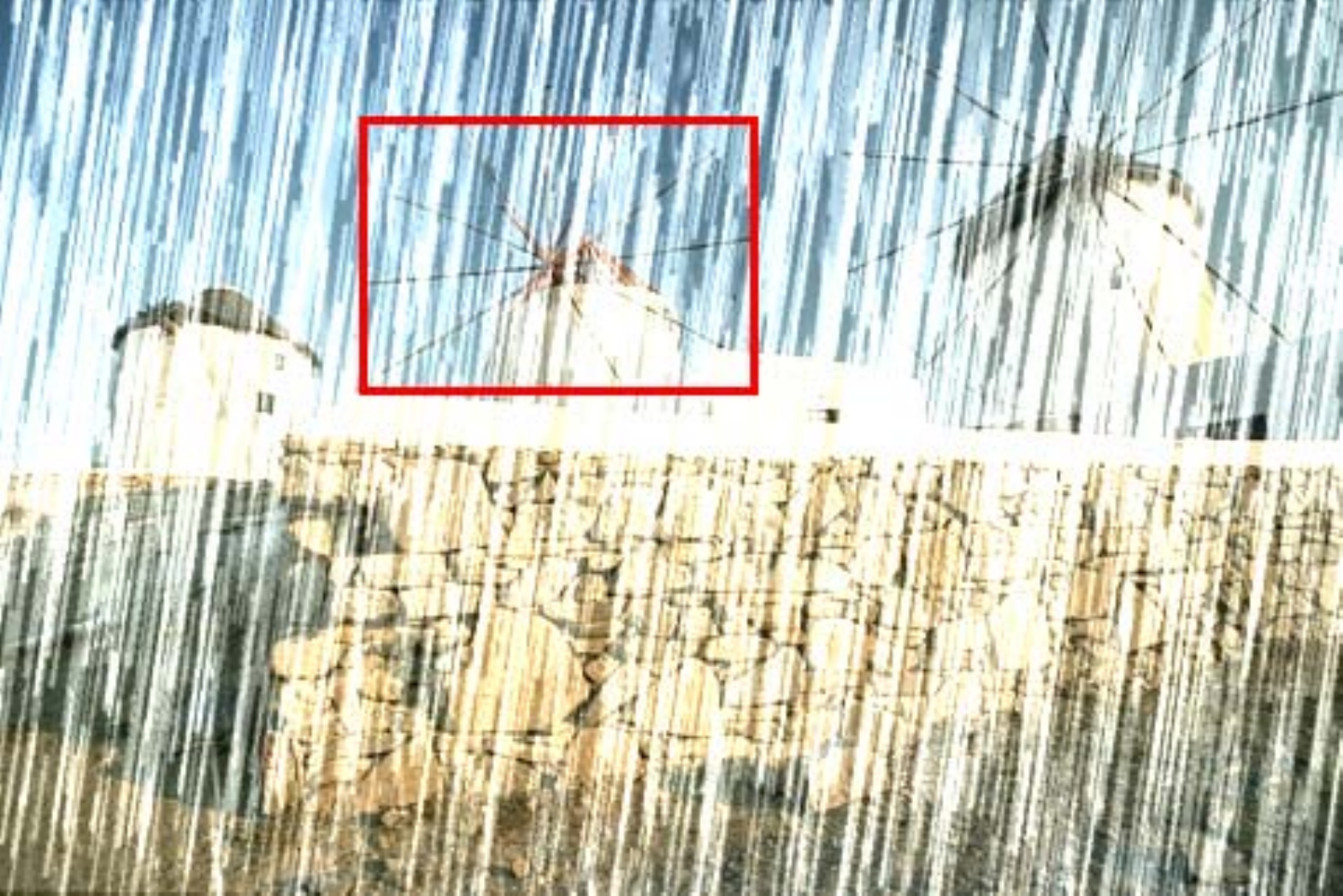} &\hspace{-4mm}
\includegraphics[width = 0.135\linewidth]{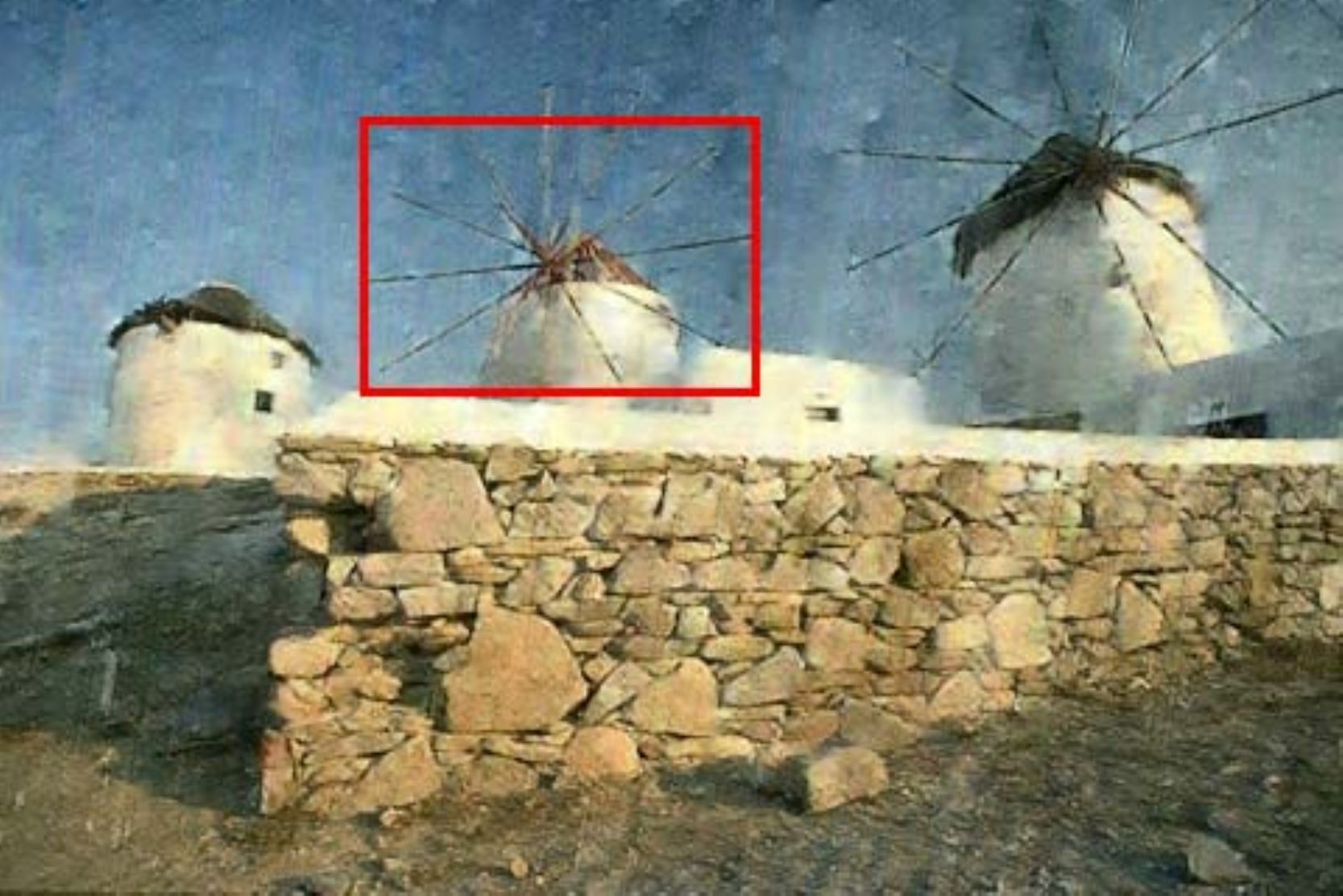} &\hspace{-4mm}
\includegraphics[width = 0.135\linewidth]{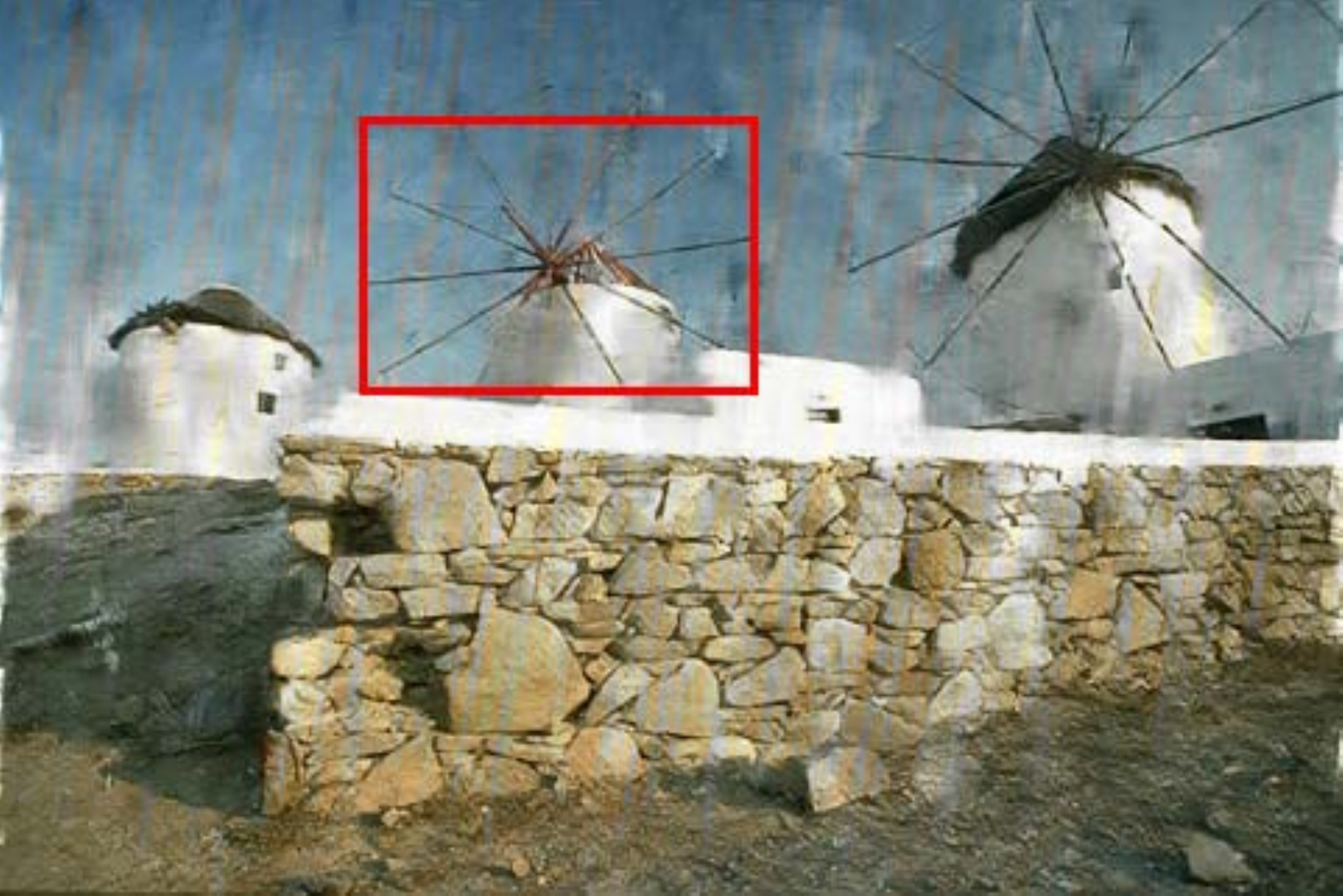} &\hspace{-4mm}
\includegraphics[width = 0.135\linewidth]{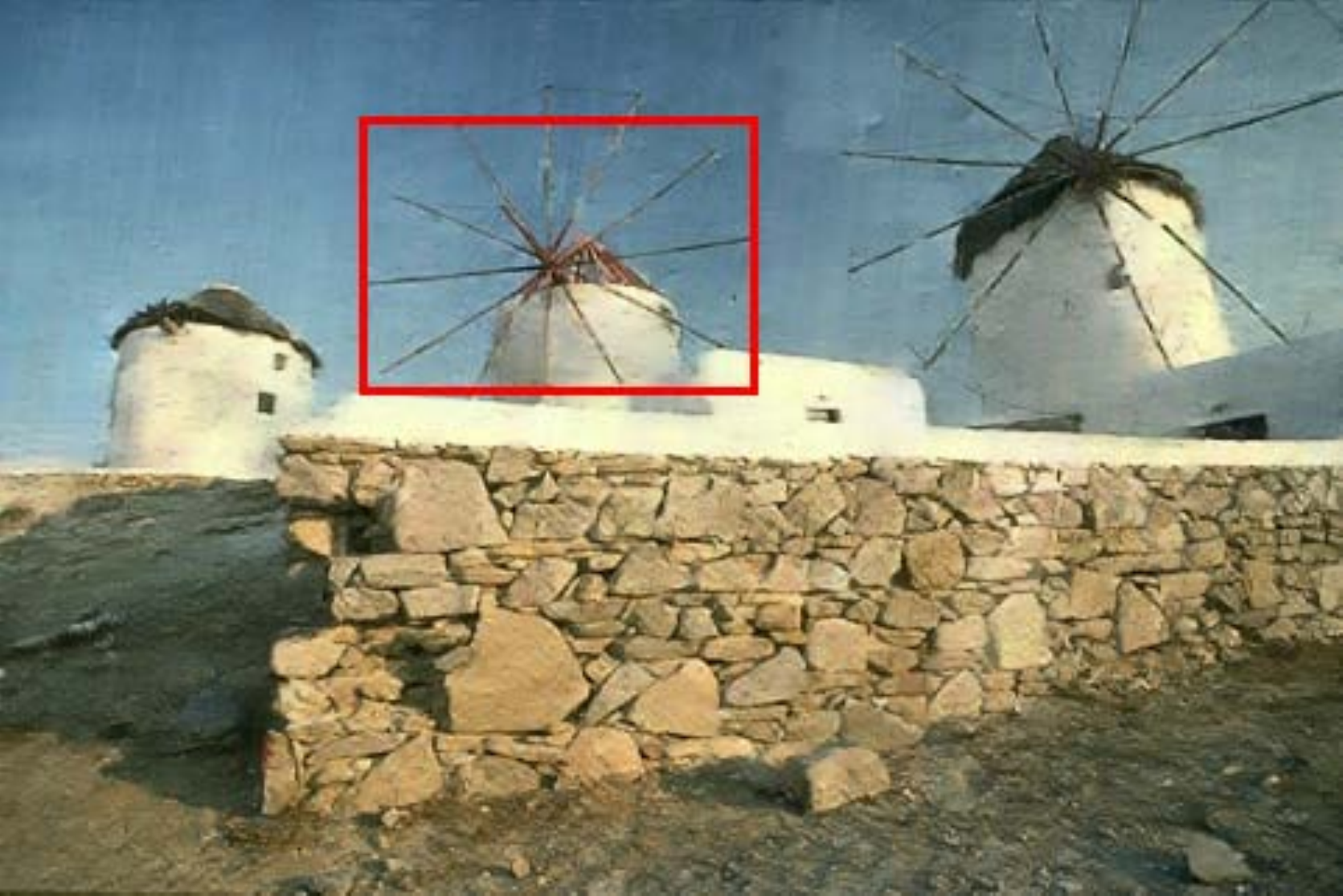} &\hspace{-4mm}
\includegraphics[width = 0.135\linewidth]{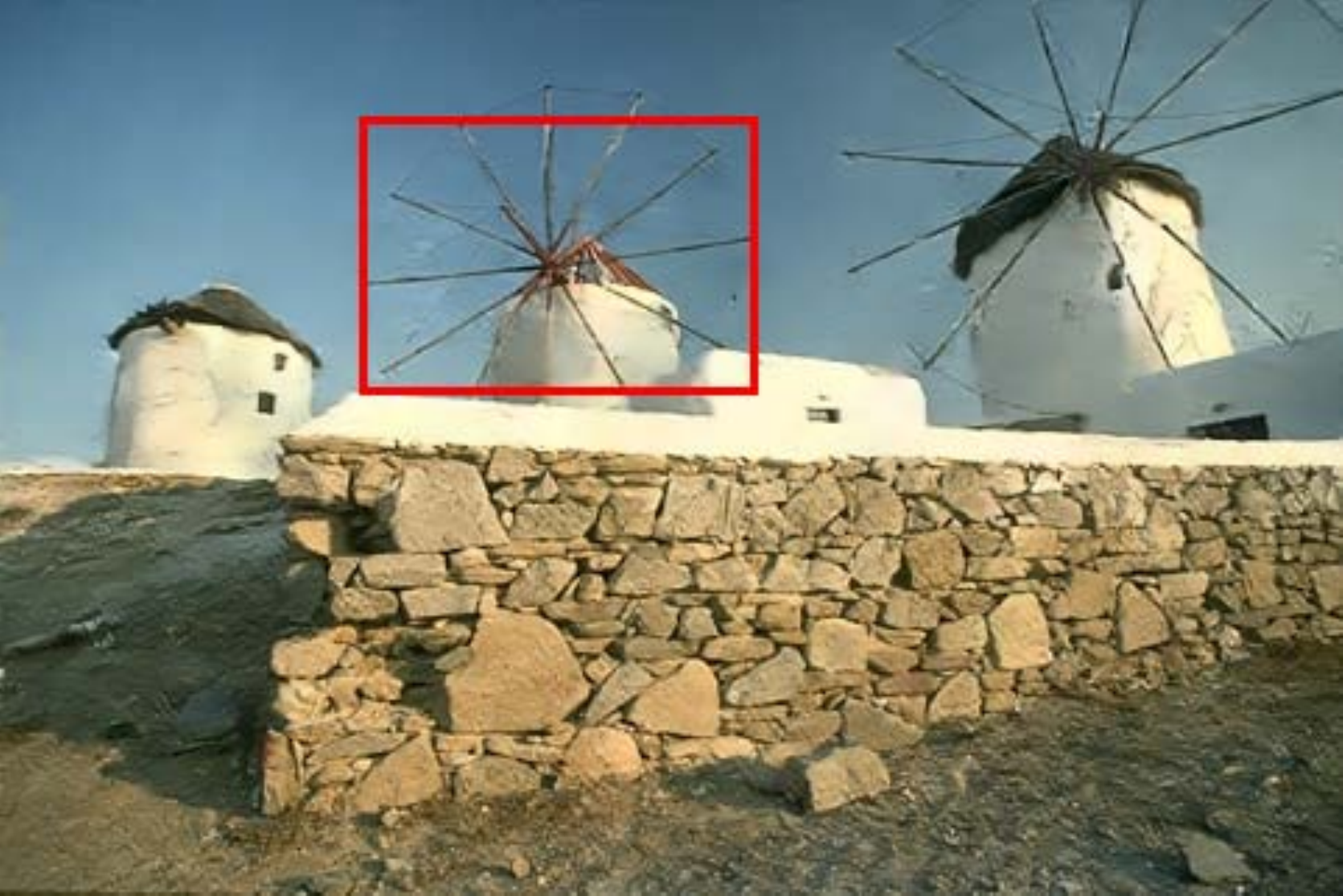} &\hspace{-4mm}
\includegraphics[width = 0.135\linewidth]{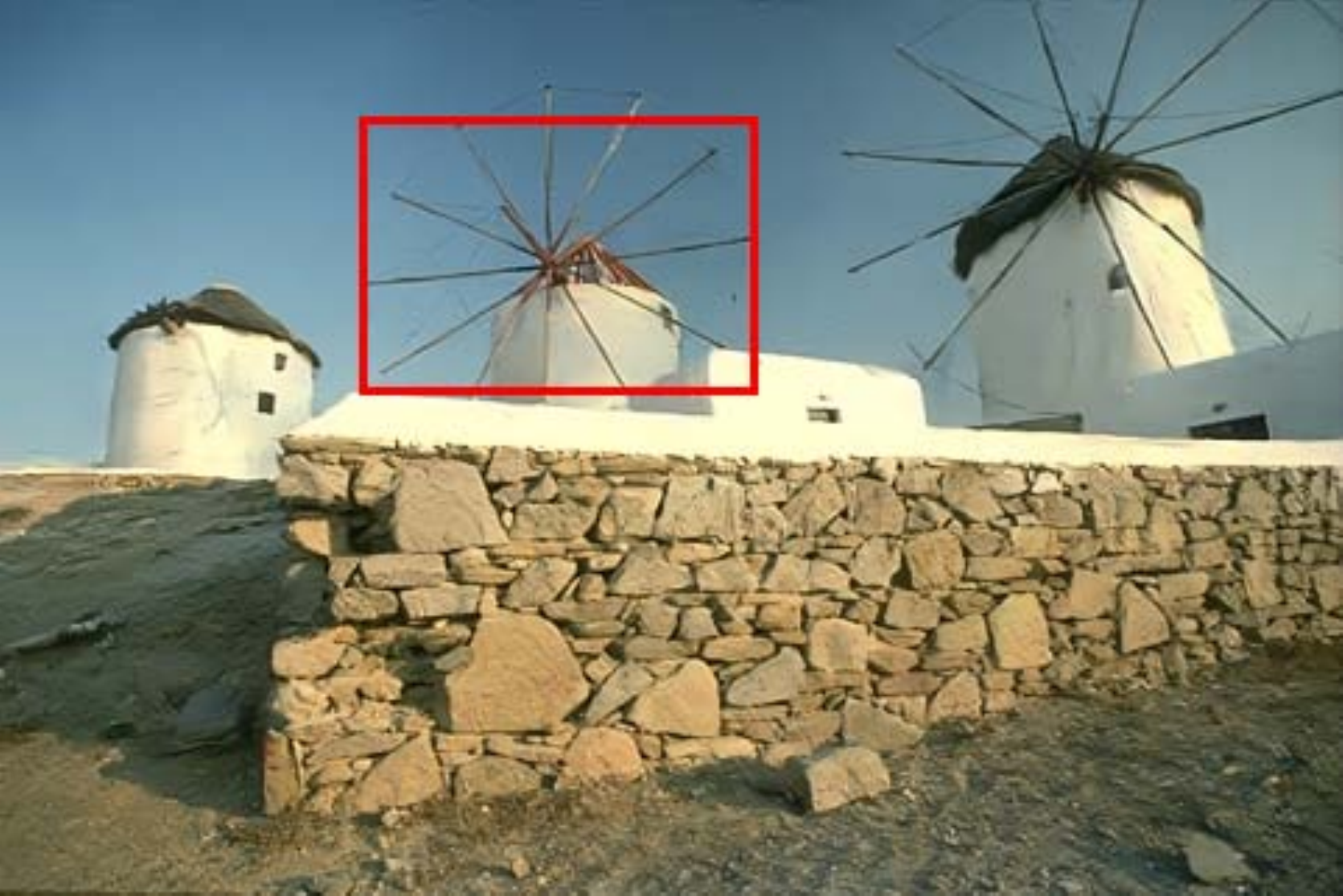} &\hspace{-4mm}
\includegraphics[width = 0.135\linewidth]{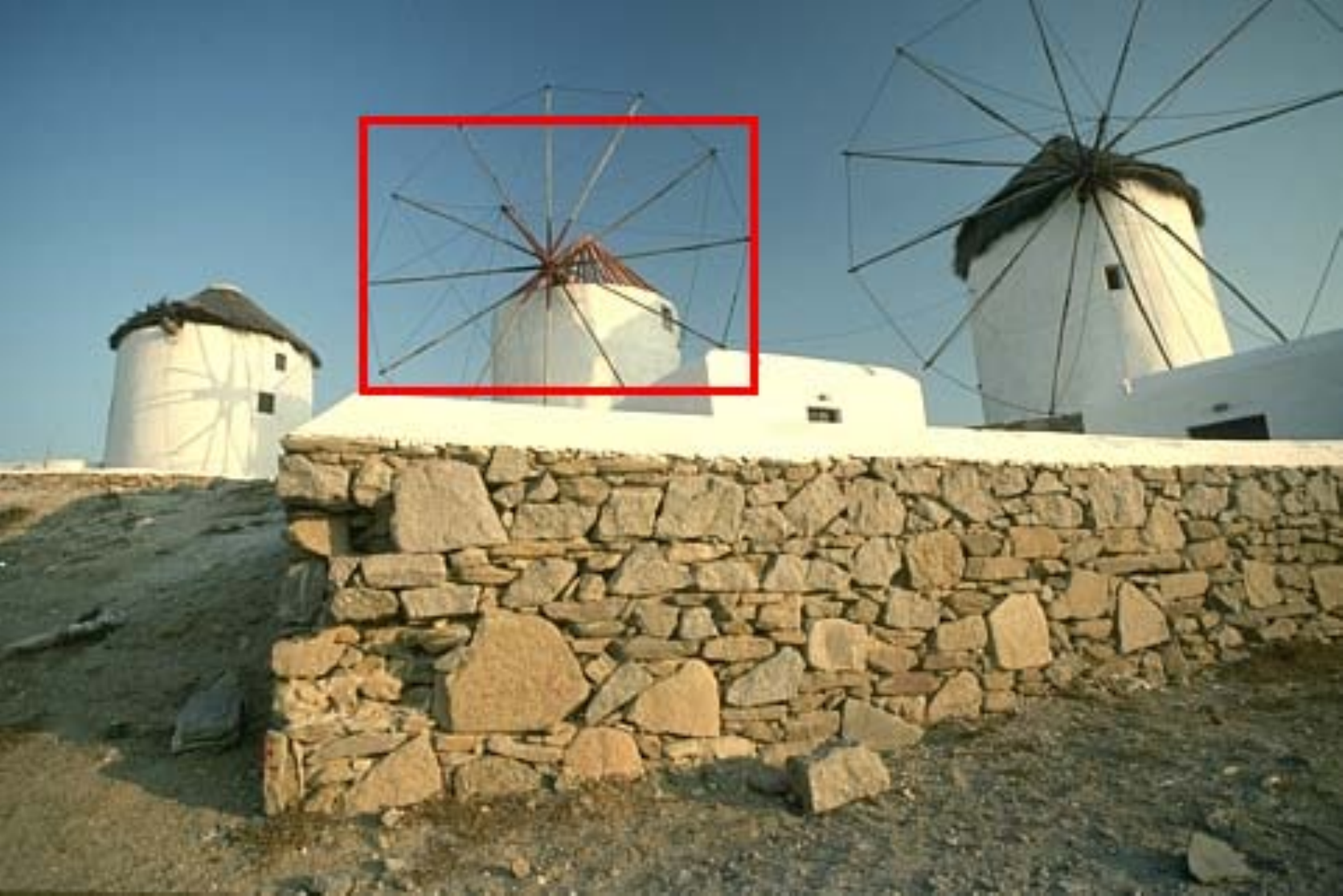}
\\
PSNR/SSIM &\hspace{-4mm}  22.89/0.71    &\hspace{-4mm}    20.86/0.69 &\hspace{-4mm} 24.73/0.72 &\hspace{-4mm}  26.20/0.88 &\hspace{-4mm}  \textbf{27.49/0.89} &\hspace{-4mm}  Inf/1
\\
\includegraphics[width = 0.135\linewidth]{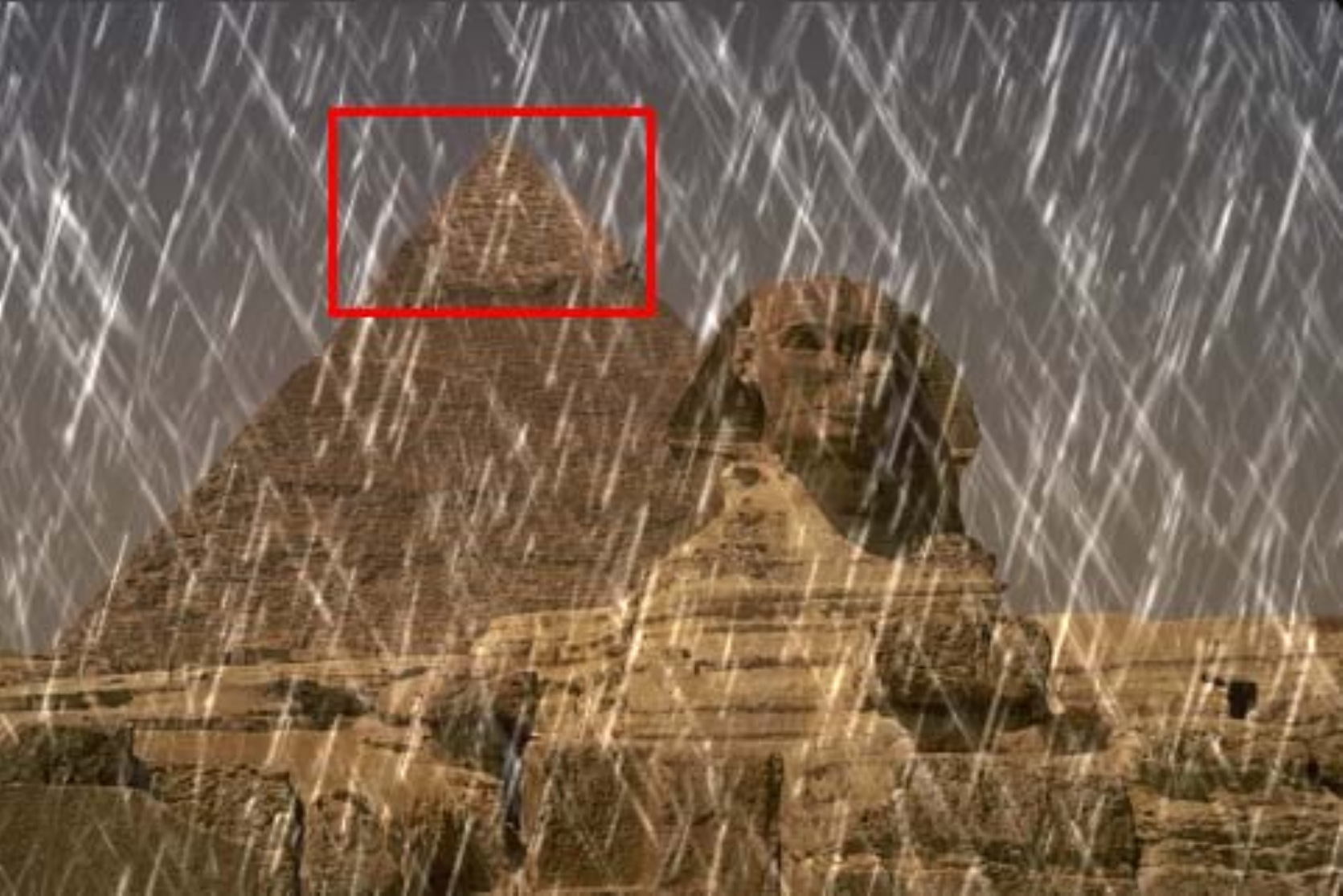} &\hspace{-4mm}
\includegraphics[width = 0.135\linewidth]{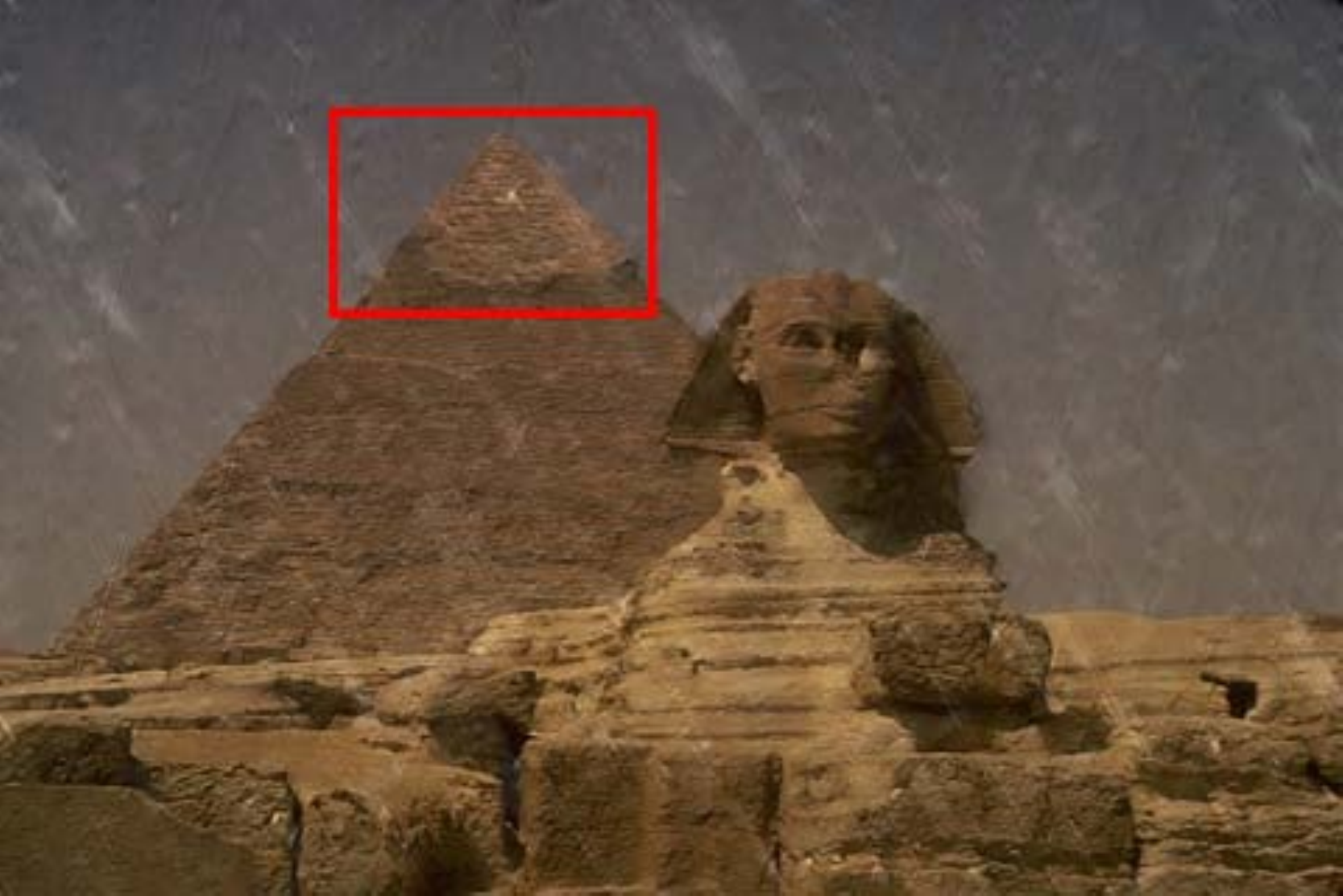} &\hspace{-4mm}
\includegraphics[width = 0.135\linewidth]{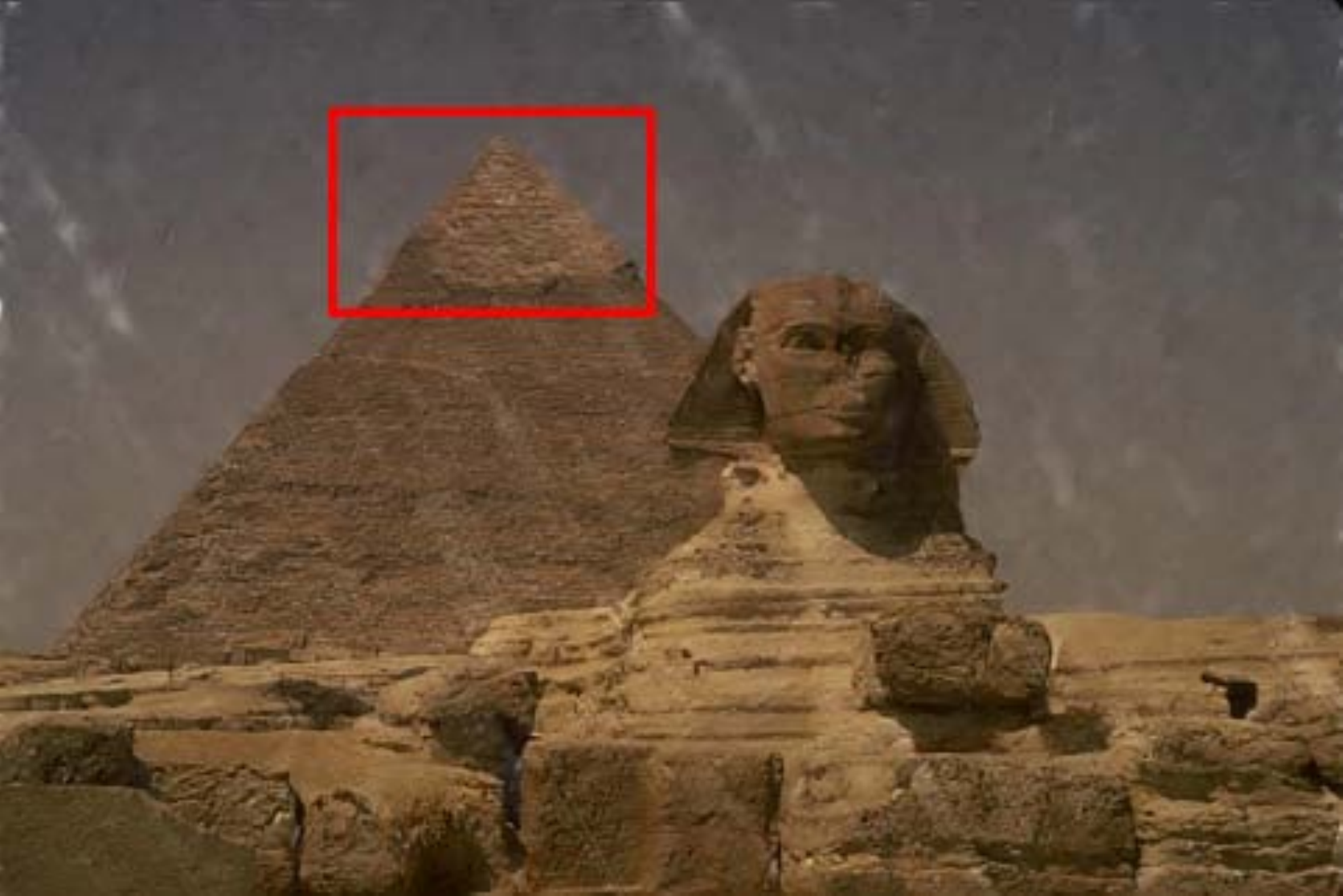} &\hspace{-4mm}
\includegraphics[width = 0.135\linewidth]{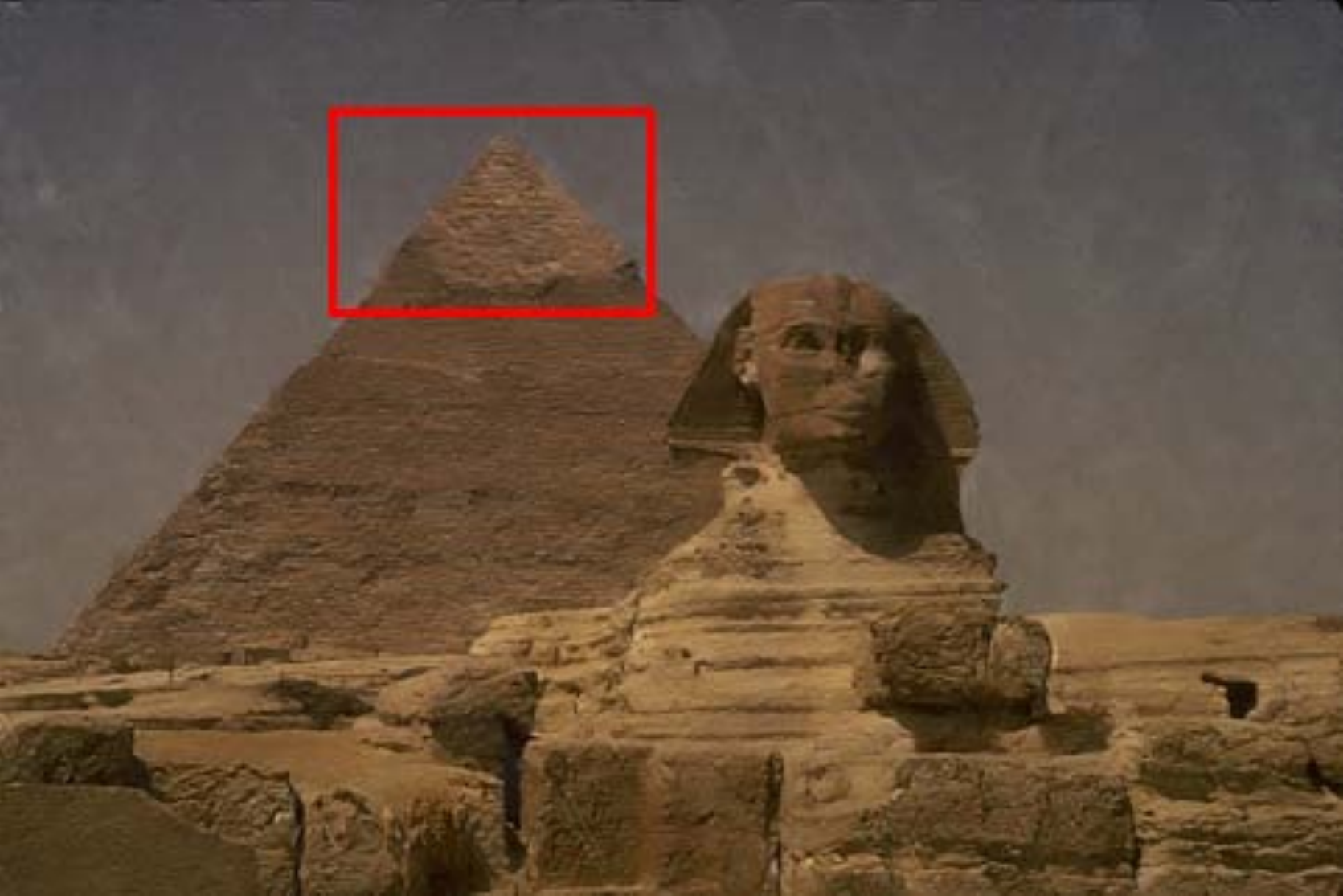} &\hspace{-4mm}
\includegraphics[width = 0.135\linewidth]{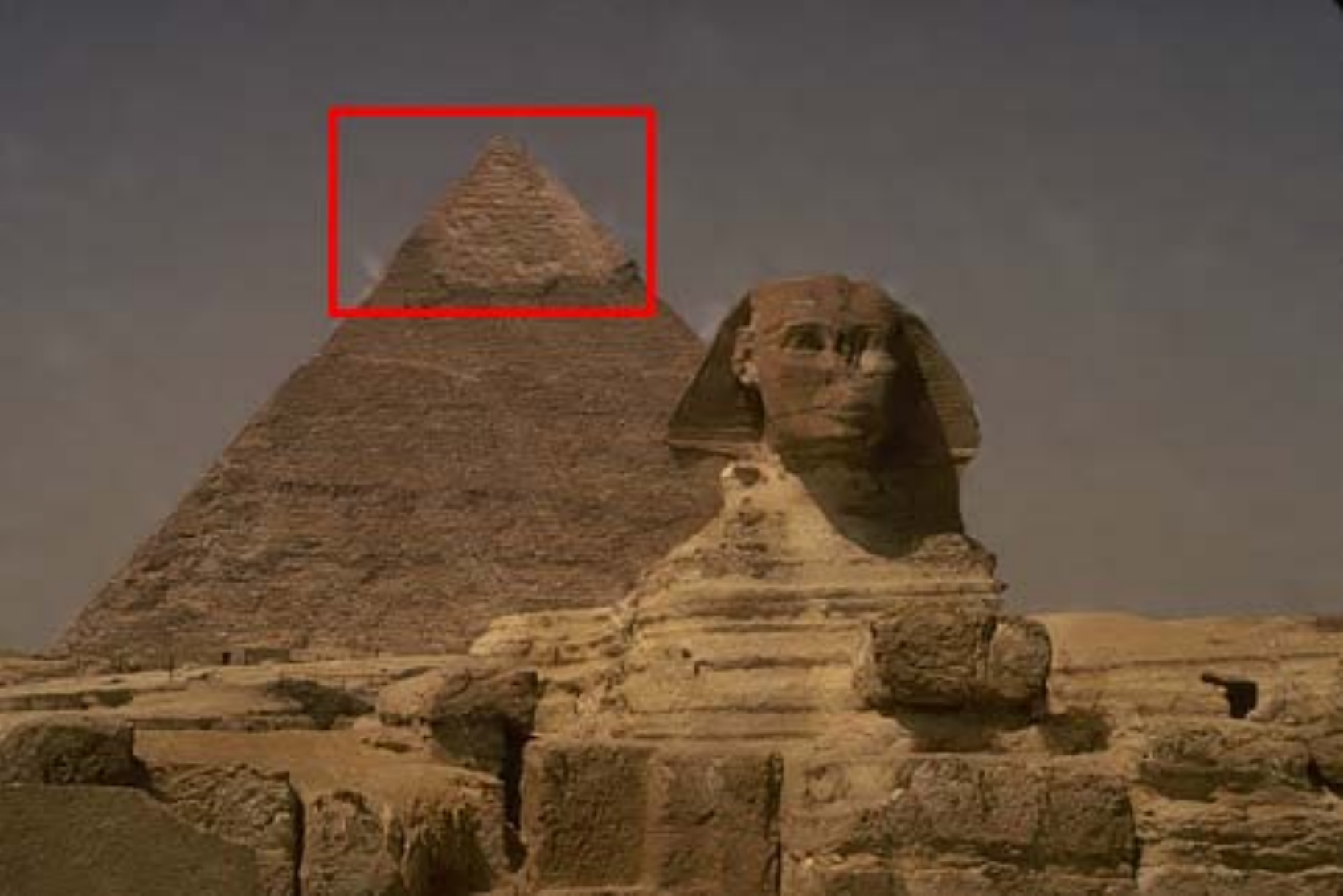} &\hspace{-4mm}
\includegraphics[width = 0.135\linewidth]{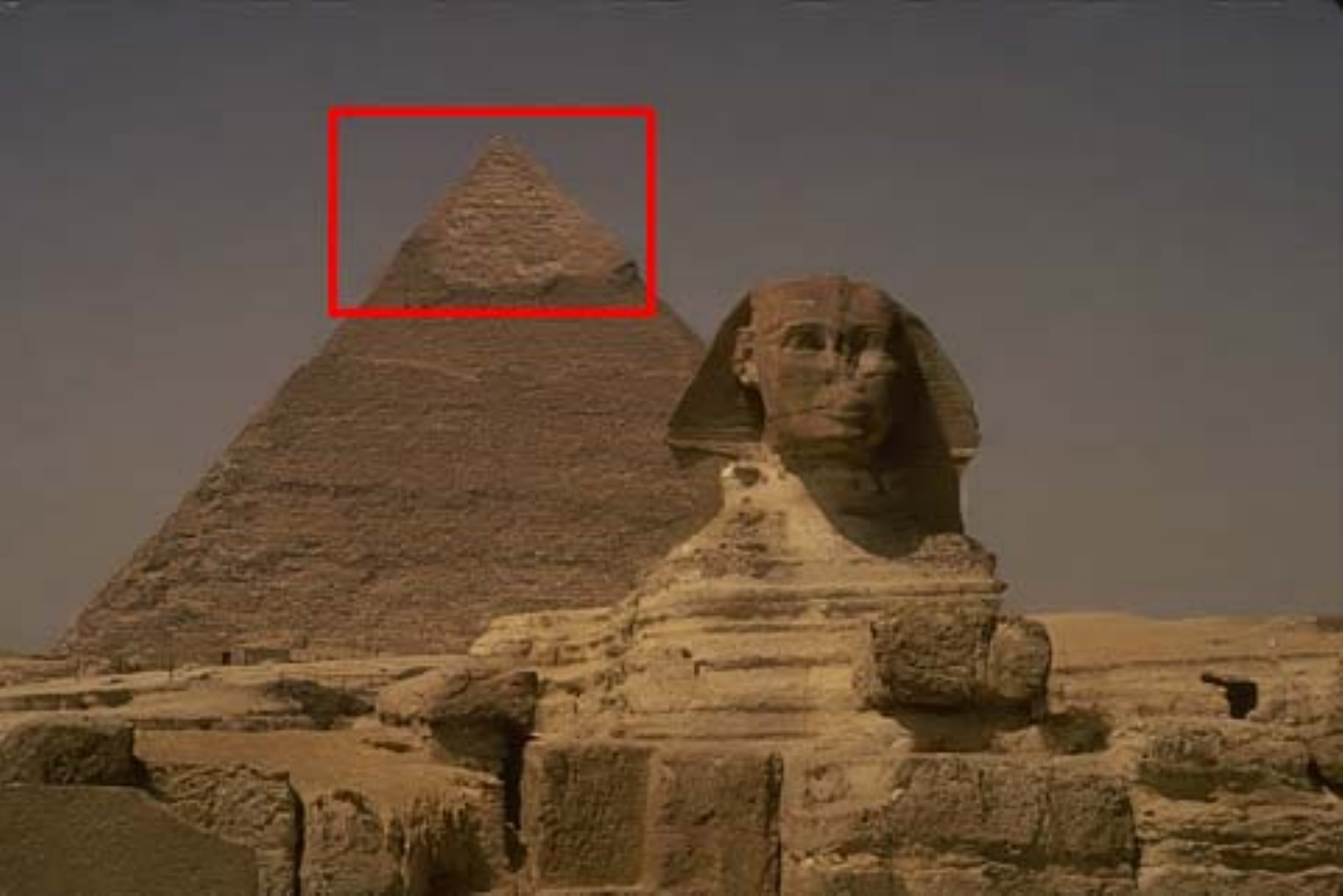} &\hspace{-4mm}
\includegraphics[width = 0.135\linewidth]{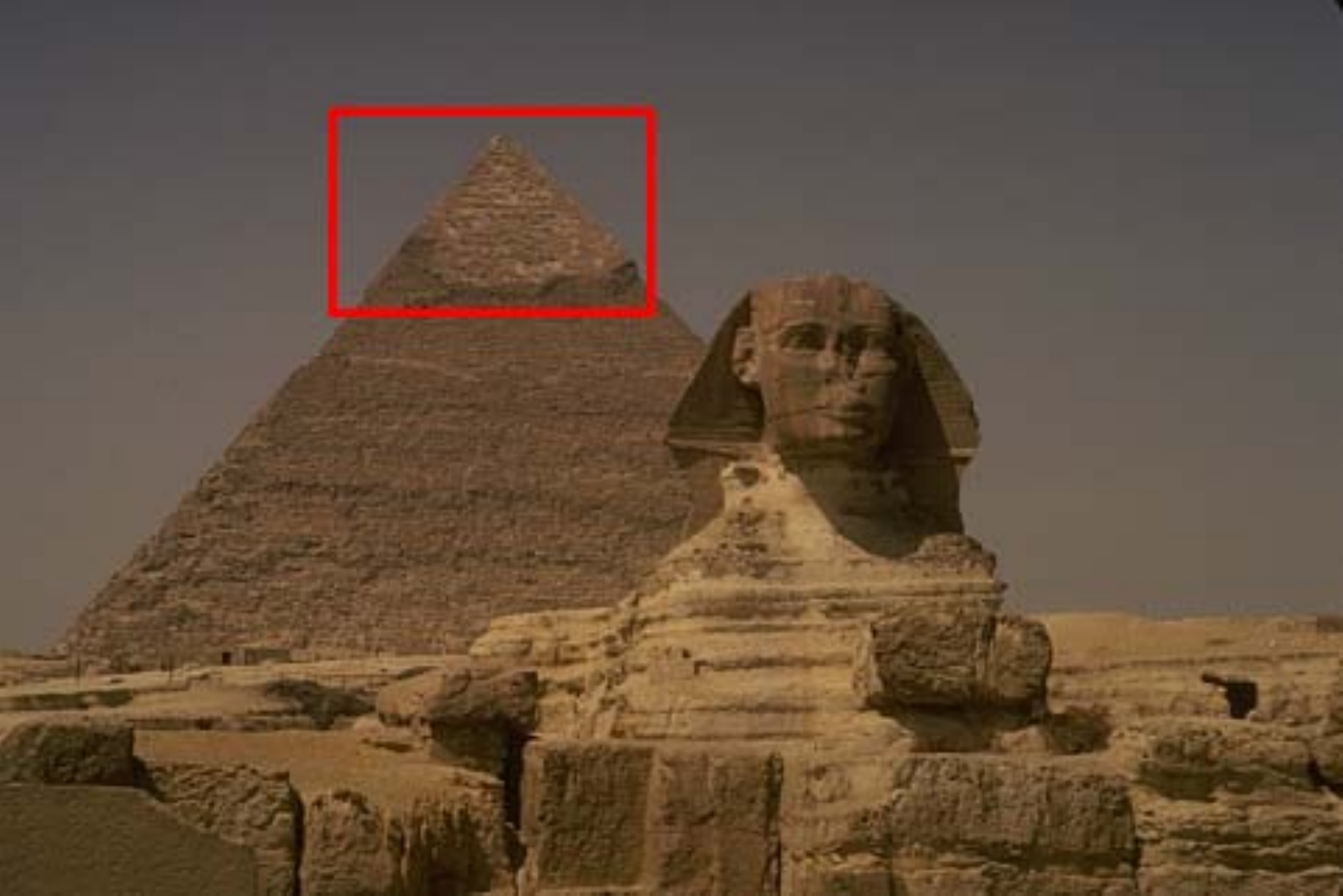}
\\
PSNR/SSIM&\hspace{-4mm}   30.80/0.88   &\hspace{-4mm}   31.20/0.91  &\hspace{-4mm} 32.74/0.87 &\hspace{-4mm}  33.81/0.95 &\hspace{-4mm}  \textbf{35.12/0.96} &\hspace{-4mm}  Inf/1
\\
(a) Input &\hspace{-4mm} (b) DDN &\hspace{-4mm} (c) JORDER &\hspace{-4mm} (d) RESCAN &\hspace{-4mm} (e) PReNet &\hspace{-4mm} (f)~\textbf{Ours} &\hspace{-4mm} (g) GT
\\
\end{tabular}
\end{center}
\vspace{-4mm}
\caption{Visual and quantitative comparisons of three synthetic examples.
Obviously, the proposed method performs better than the other four deep learning-based methods, especially the region in masked box.
Our results shown in (f) have the highest PSNR and SSIM values and are the cleanest.
}
\label{fig: visual examples on synthetic images}
\vspace{-3mm}
\end{figure*}
In this section, we conduct a number of deraining experiments on three synthetic datasets and real-world datasets compared with six state-of-the-art deraining methods, including discriminative sparse coding (DSC)~\cite{derain_dsc_luo}, layer priors (LP)~\cite{derain_lp_li}, deep detail network (DDN)~\cite{derain_ddn_fu}, the recurrent version of joint rain detection and removal (JORDER)~\cite{derain_jorder_yang}, RESCAN~\cite{derain_rescan_li} and PReNet~\cite{derain_prenet_Ren_2019_CVPR}.

\subsection{Experiment settings}

\label{sec: Experiment settings}
{\flushleft \bf Synthetic Datasets}.
We carry out experiments to evaluate the performance of our method on three synthetic datasets: Rain100H, Rain100L, and Rain1200, which all have various rain streaks with different sizes, shapes, and directions.
There are 1800 image pairs for training and 200 image pairs for testing in Rain100H and Rain100L.
In Rain1200, 12000 images are for training and 1200 images for testing.
We choose Rain100H as our analysis dataset.
\vspace{-1mm}
{\flushleft \bf Real-world Testing Images}.
We also evaluate the performance of our method on real-world images, which are provided by Zhang et al.~\cite{derain_cgan_zhang} and Yang et al.~\cite{derain_jorder_yang}.
In these images, they have different rain components from orientation to density.

\vspace{-1mm}
{\flushleft \bf Training Settings}.
In the training process, we randomly crop each training image pairs to $160\times160$ patch pairs.
The batch size is chosen as 64.
For each convolution layer, we use leaky-ReLU with $\alpha = 0.2$ as the activation function except for the last layer.
We use ADAM algorithm~\cite{adam} to optimize our network.
The initial learning rate is $5\times 10^{-4}$, and is updated twice by a rate of $1/10$ at 1200 and 1600 epochs and the total epoch is 2000.
$\alpha, \beta$ and $\gamma$ are set as 0.5, 0.5 and 0.001, respectively.
Our entire network is trained on 8 Nvidia GTX 1080Ti GPUs based on PyTorch.
\vspace{-1mm}
{\flushleft \bf Evaluation Criterions}.
We use peak signal to noise ratio (PSNR) and structure similarity (SSIM) to evaluate the quality of the recovered results in comparison with ground truth images.
PSNR and SSIM are only computed for synthetic datasets, because not only the estimated rain-free images are needed, but also corresponding ground truth images during the computing process.
For real-world images, they can only be evaluated by visual comparisons.

\subsection{Results on synthetic datasets}
\label{sec: Results on synthetic datasets}
Tab.~\ref{tab:synthetic datasets} shows quantitatively comparative results between our method and six state-of-the-art deraining methods on Rain100H, Rain100L and Rain1200.
There are two conventional methods: DSC~\cite{derain_dsc_luo} (ICCV15) and LP~\cite{derain_lp_li} (CVPR16), and four deep learning-based methods: DDN~\cite{derain_ddn_fu} (CVPR17), JORDER~\cite{derain_jorder_yang} (CVPR17), RESCAN~\cite{derain_rescan_li} (ECCV18) and PReNet~\cite{derain_prenet_Ren_2019_CVPR} (CVPR19).
As we can see that our proposed method outperforms these state-of-the-art approaches on the three datasets.

We also show several challenging synthetic examples for visual comparisons in Fig.~\ref{fig: visual examples on synthetic images}.
As the prior based methods are obviously worse than deep learning-based methods according to Tab.~\ref{tab:synthetic datasets}, we only compare the visual performances with deep learning methods.
The first column in Fig.~\ref{fig: visual examples on synthetic images} are synthetic images that are severely degraded by rain streaks.
Fig~\ref{fig: visual examples on synthetic images}~(b) and Fig.~\ref{fig: visual examples on synthetic images}~(c) are the results of DDN~\cite{derain_ddn_fu} and JORDER~\cite{derain_jorder_yang}.
Obviously, they both fail to recover an acceptable clean image.
Fig.~\ref{fig: visual examples on synthetic images}~(d) and Fig.~\ref{fig: visual examples on synthetic images}~(e) are the results of RESCAN~\cite{derain_rescan_li} and PReNet~\cite{derain_prenet_Ren_2019_CVPR}, which have unpleasing artifacts in the masked boxes.
As shown in Fig.~\ref{fig: visual examples on synthetic images}~(f), our results generate best deraining results no matter in quantitatively or visually.

\begin{figure*}[!t]
\vspace{-5mm}
\begin{center}
\begin{tabular}{cccccc}
\includegraphics[width = 0.15\linewidth]{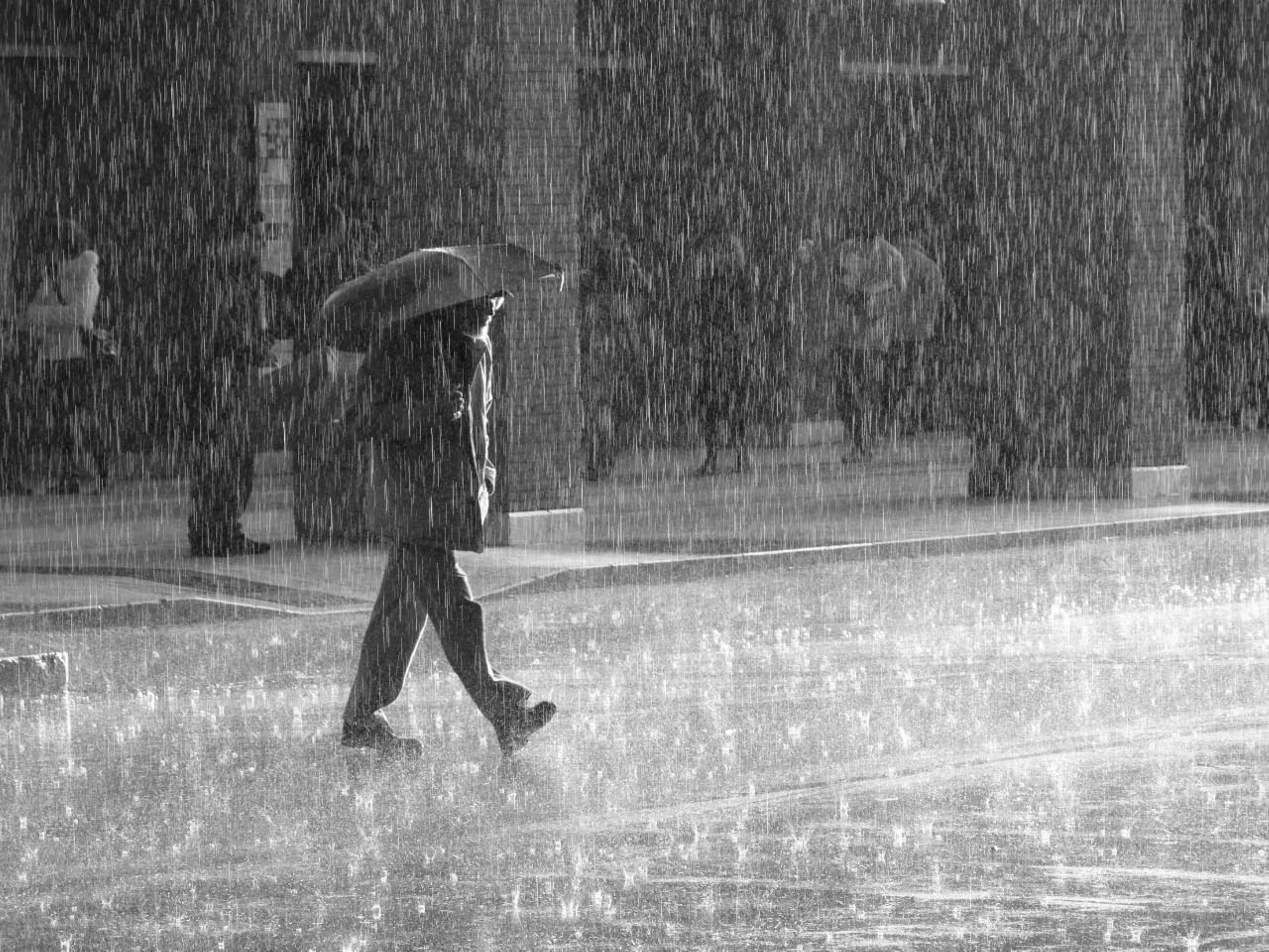} &\hspace{-4mm}
\includegraphics[width = 0.15\linewidth]{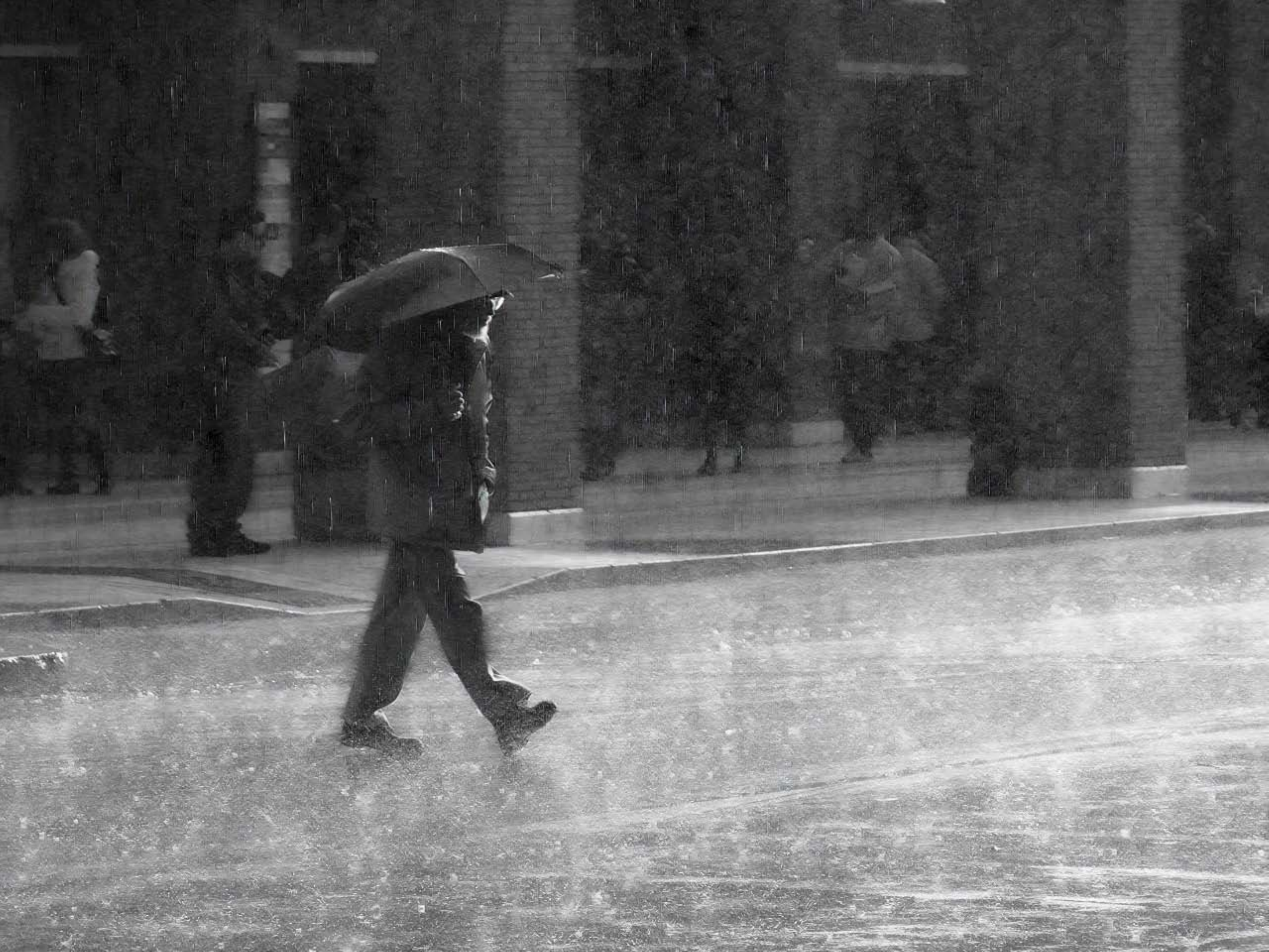} &\hspace{-4mm}
\includegraphics[width = 0.15\linewidth]{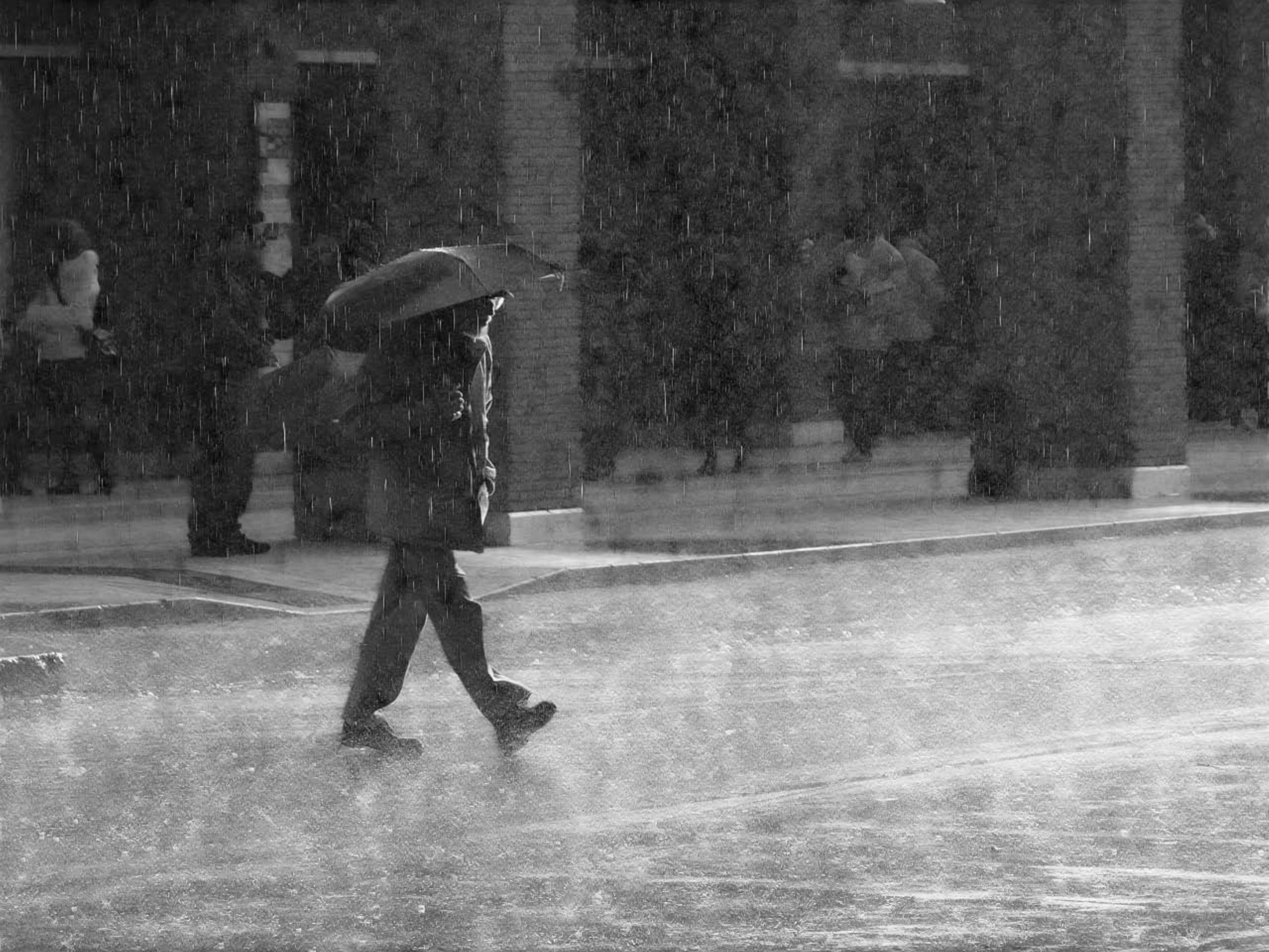} &\hspace{-4mm}
\includegraphics[width = 0.15\linewidth]{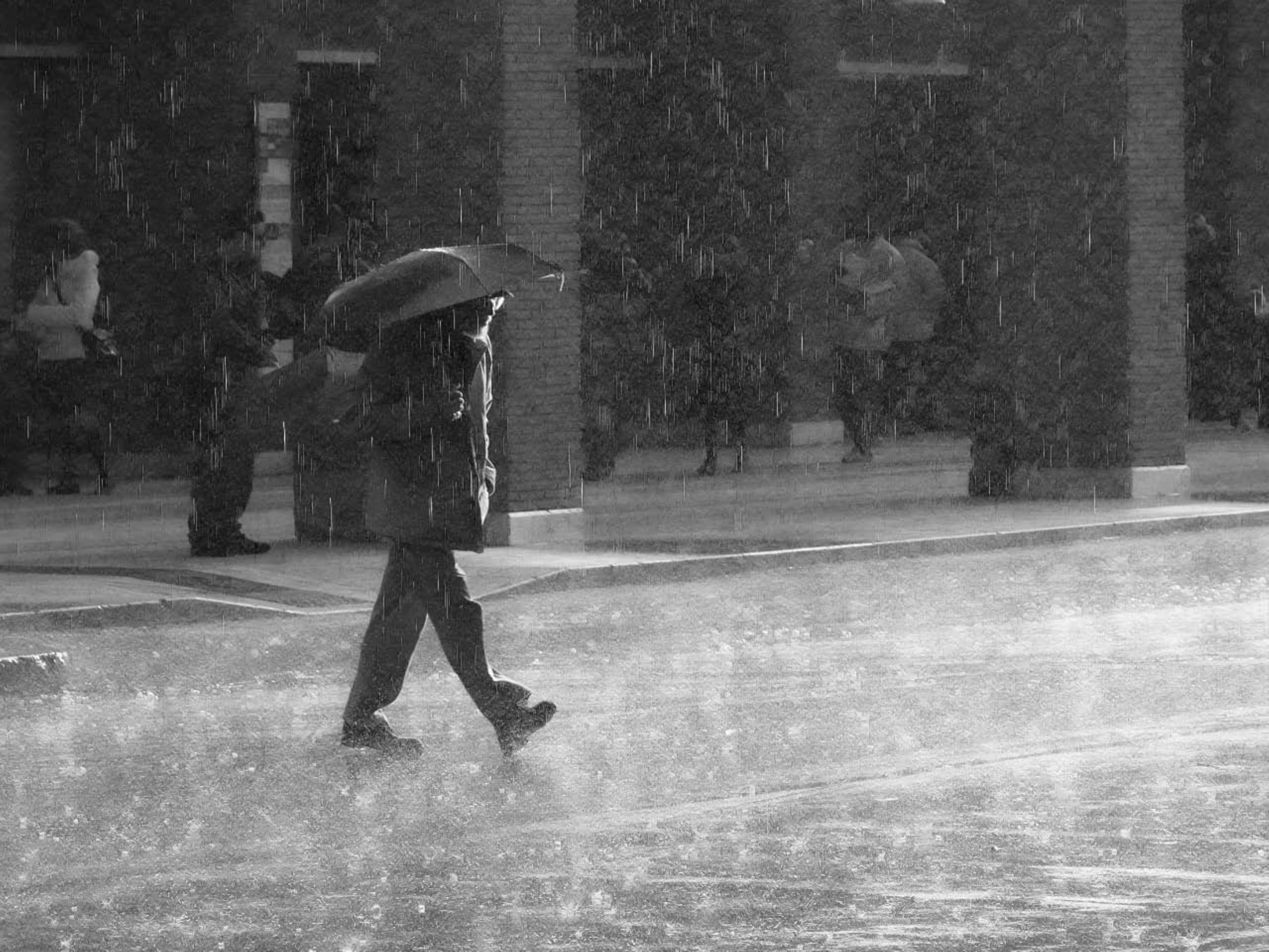} &\hspace{-4mm}
\includegraphics[width = 0.15\linewidth]{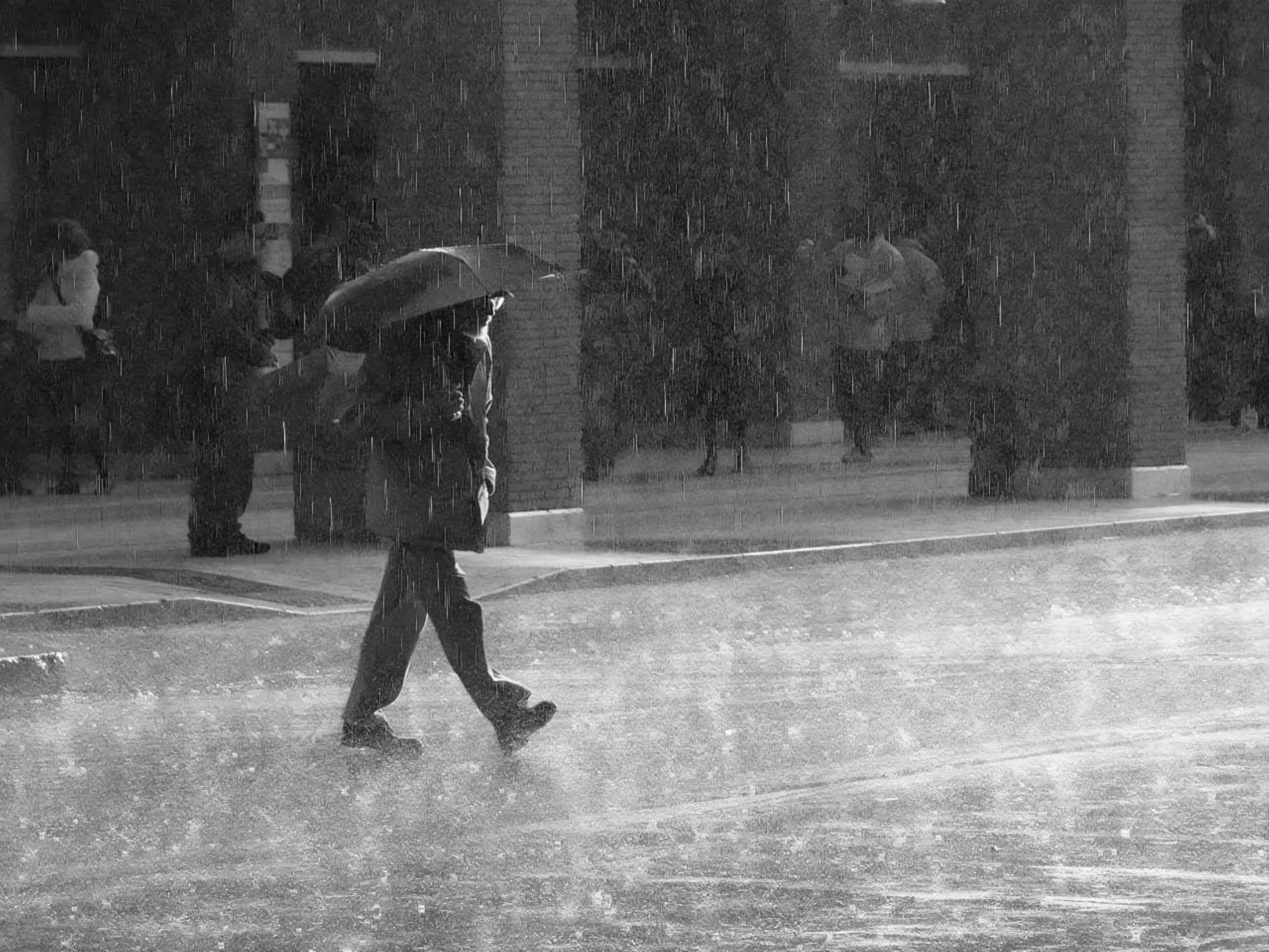} &\hspace{-4mm}
\includegraphics[width = 0.15\linewidth]{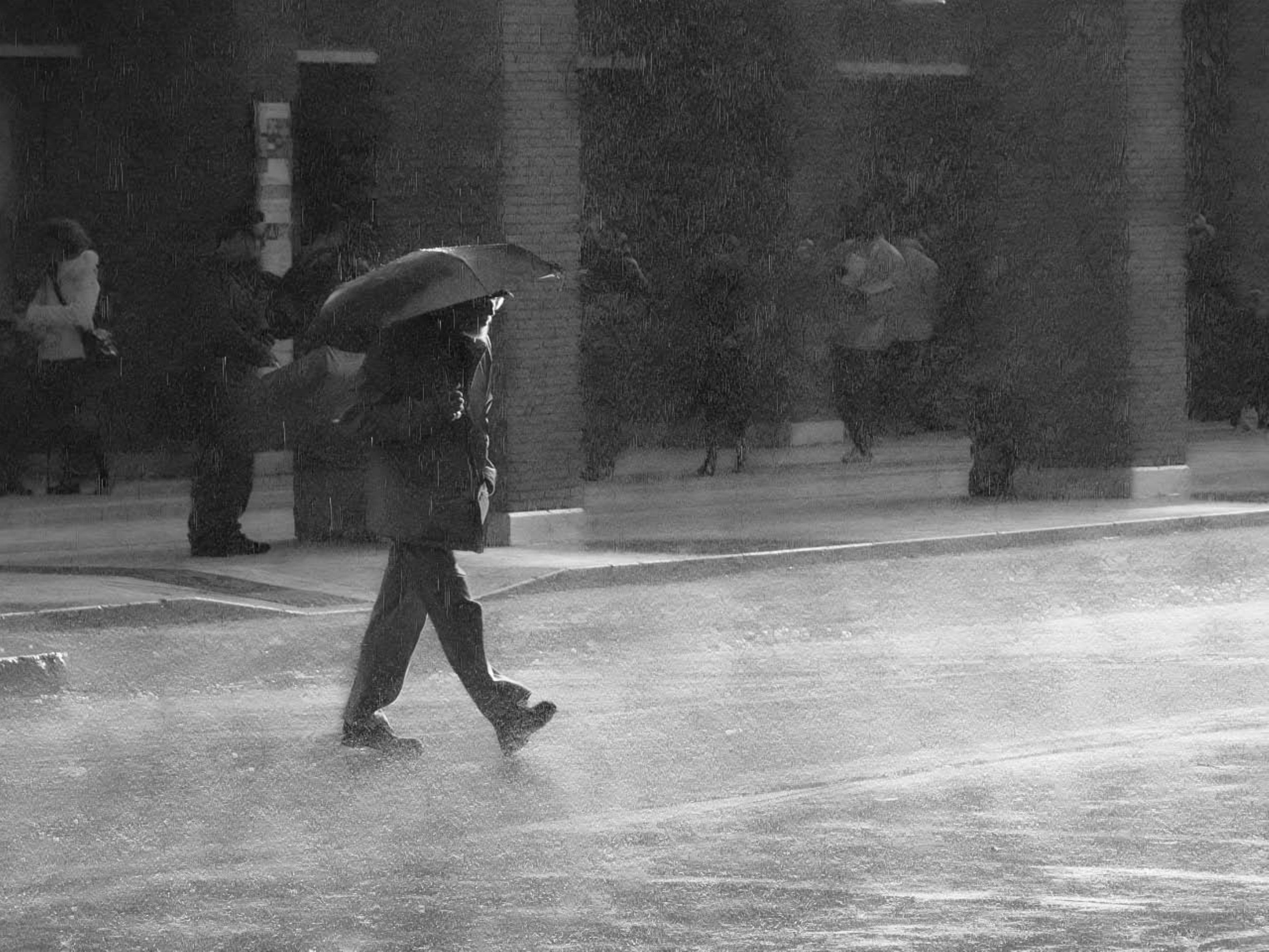}
\\
\includegraphics[width = 0.15\linewidth]{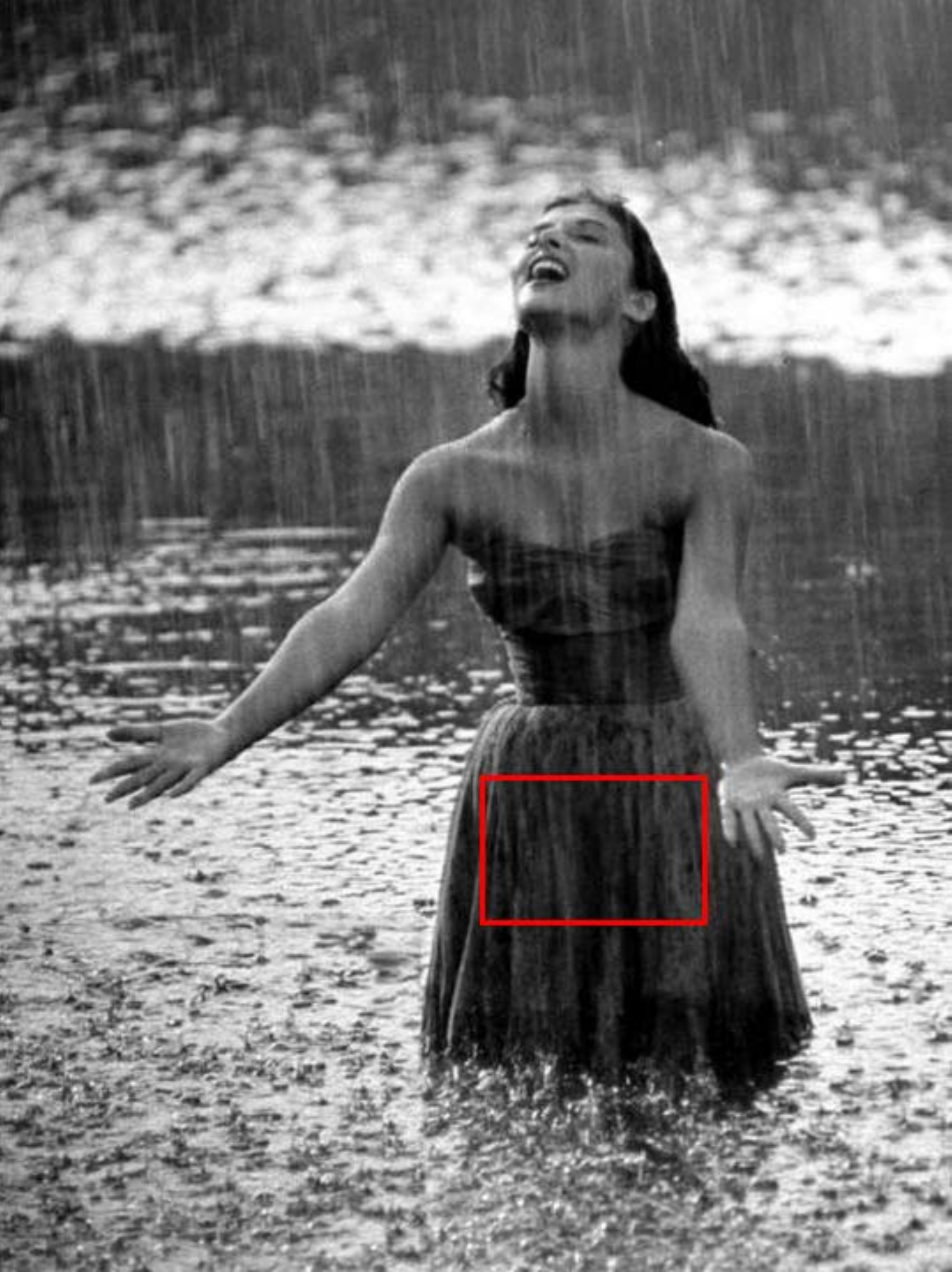} &\hspace{-4mm}
\includegraphics[width = 0.15\linewidth]{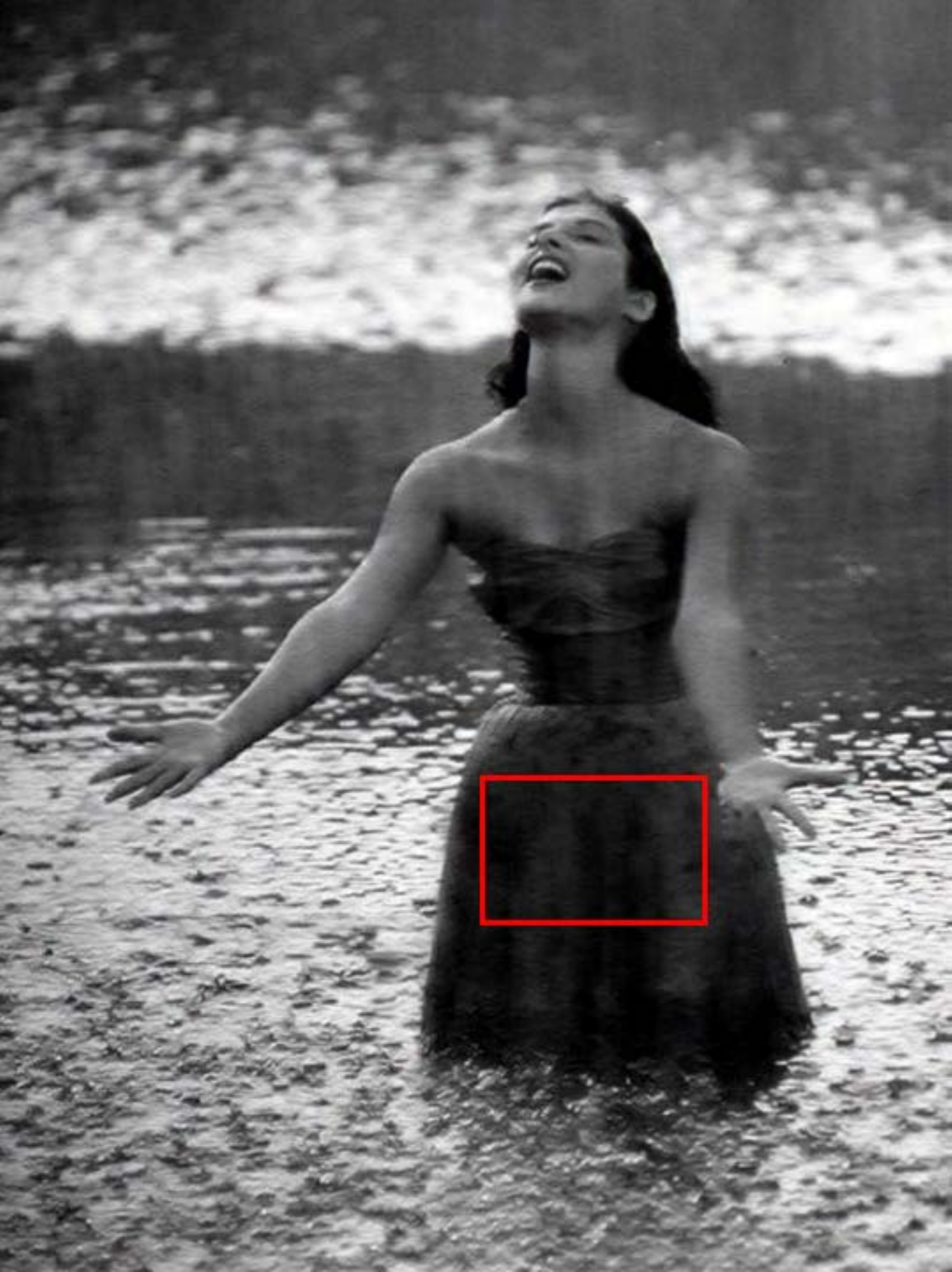} &\hspace{-4mm}
\includegraphics[width = 0.15\linewidth]{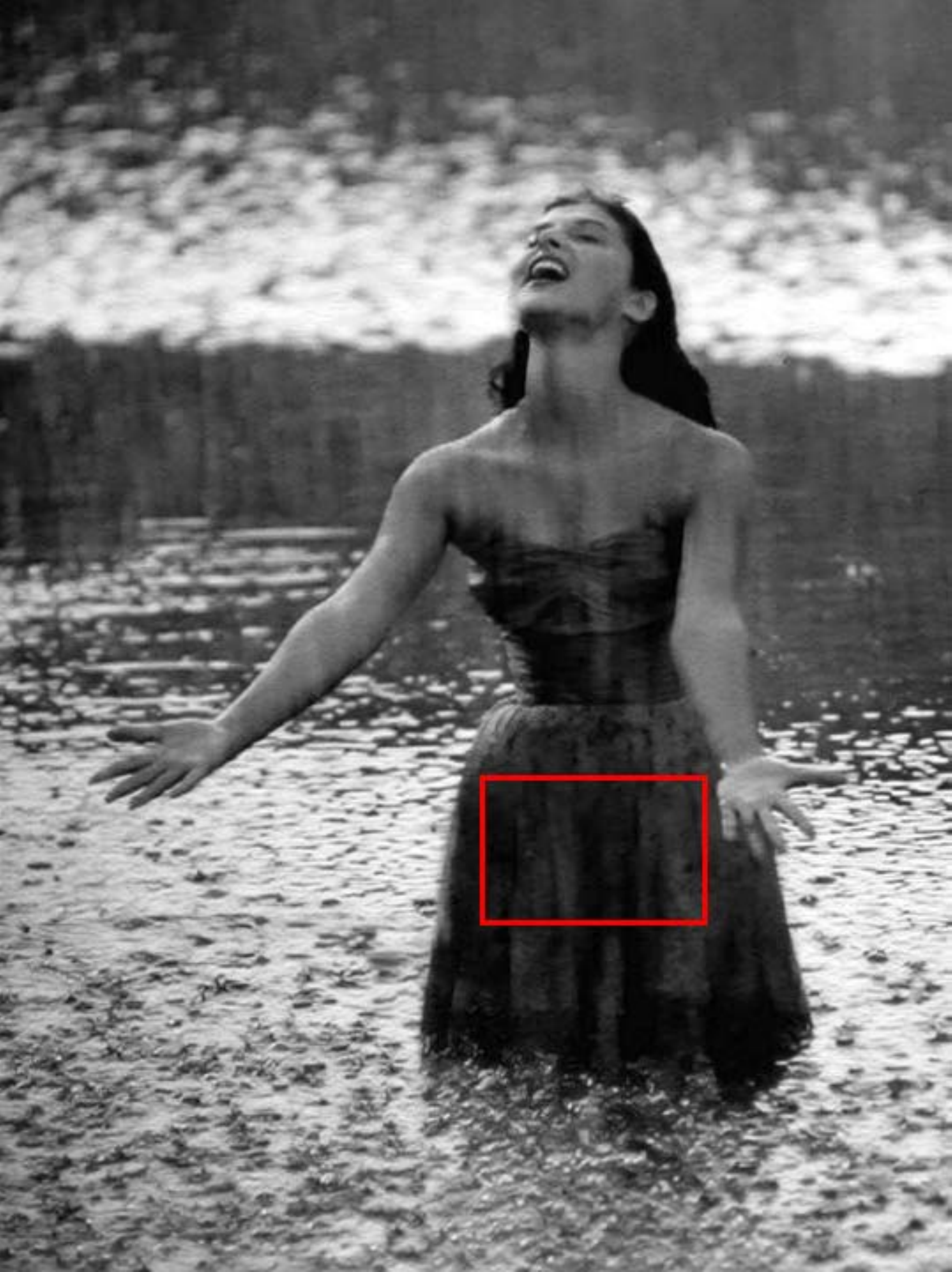} &\hspace{-4mm}
\includegraphics[width = 0.15\linewidth]{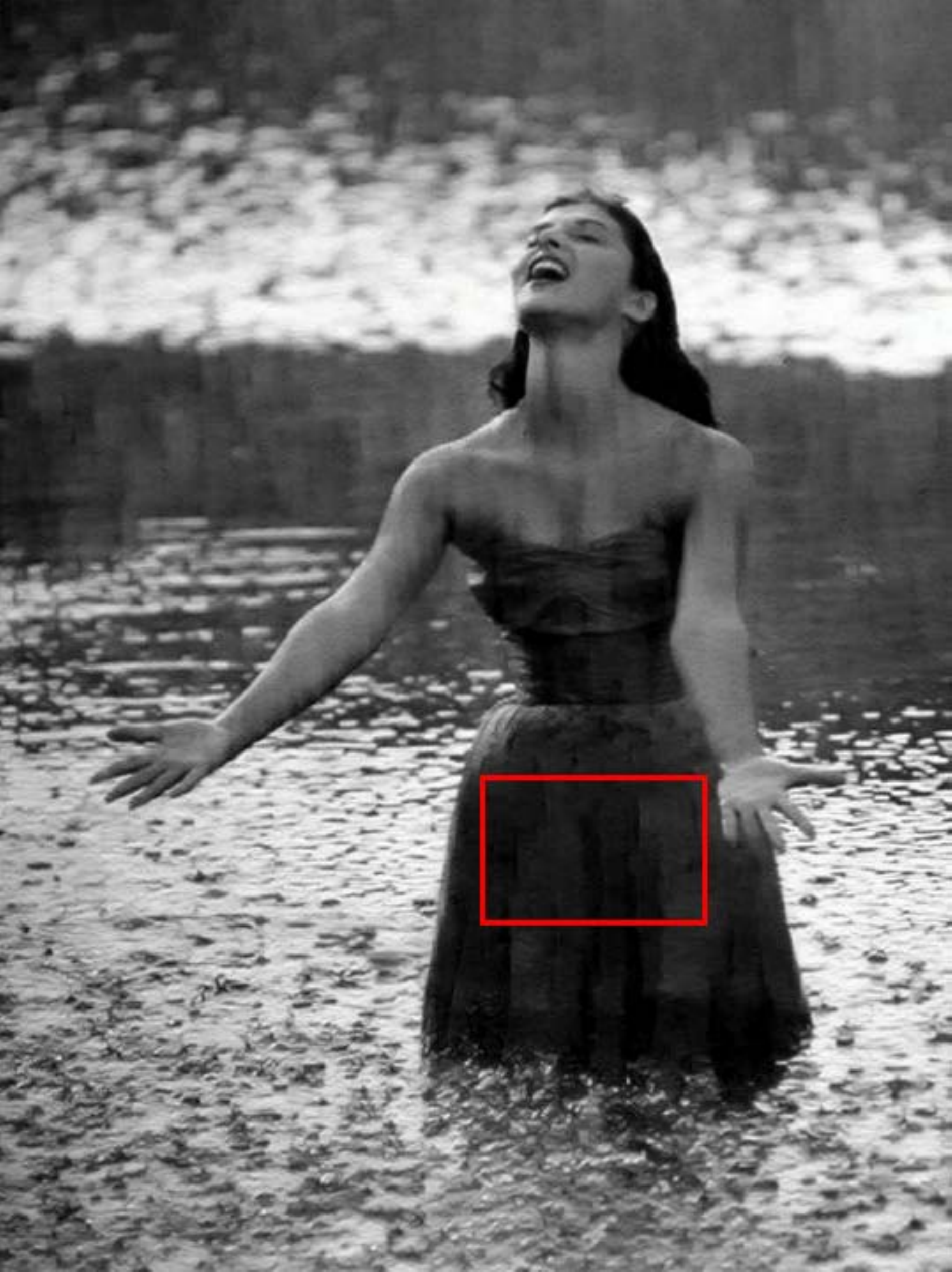} &\hspace{-4mm}
\includegraphics[width = 0.15\linewidth]{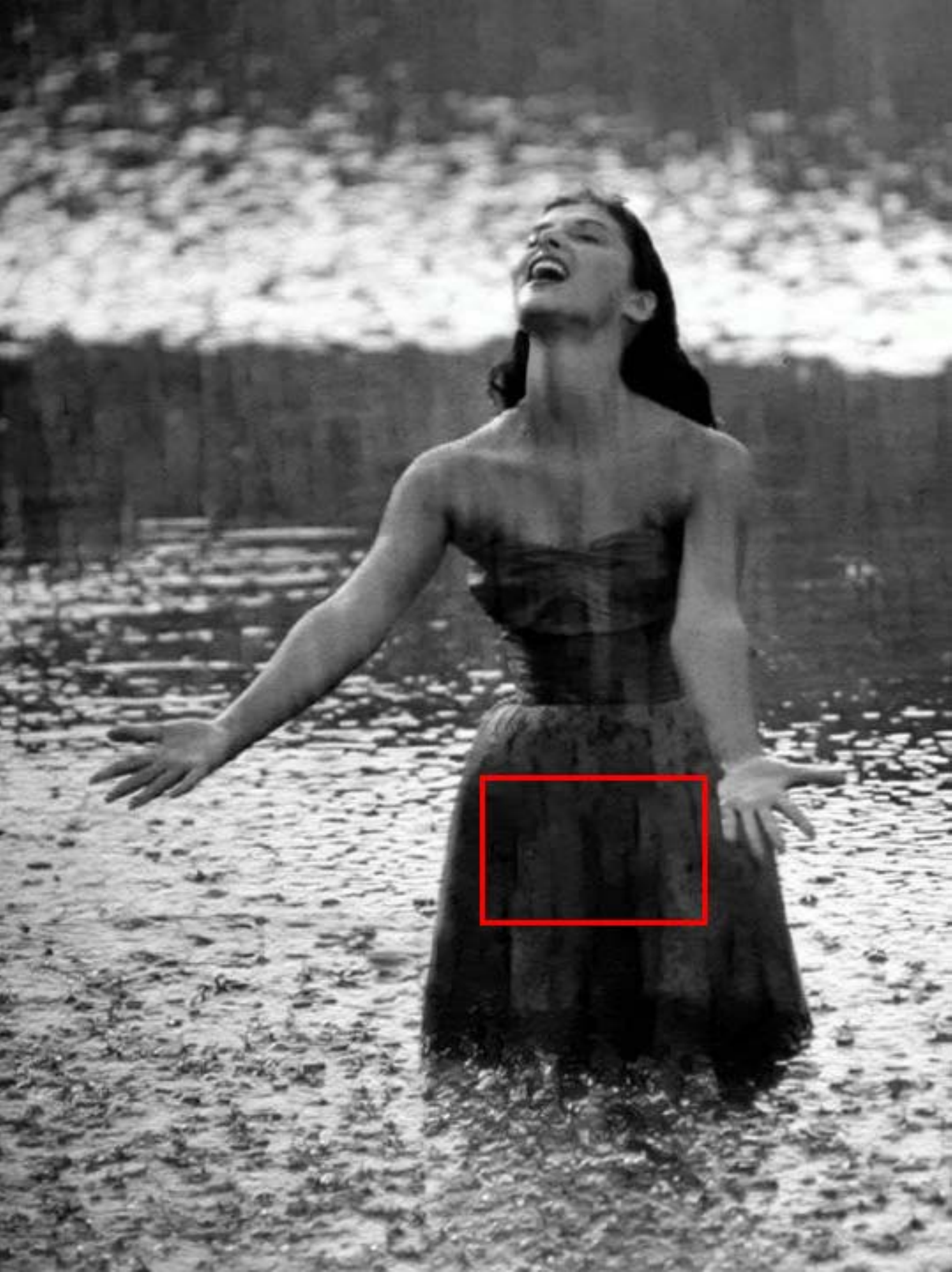} &\hspace{-4mm}
\includegraphics[width = 0.15\linewidth]{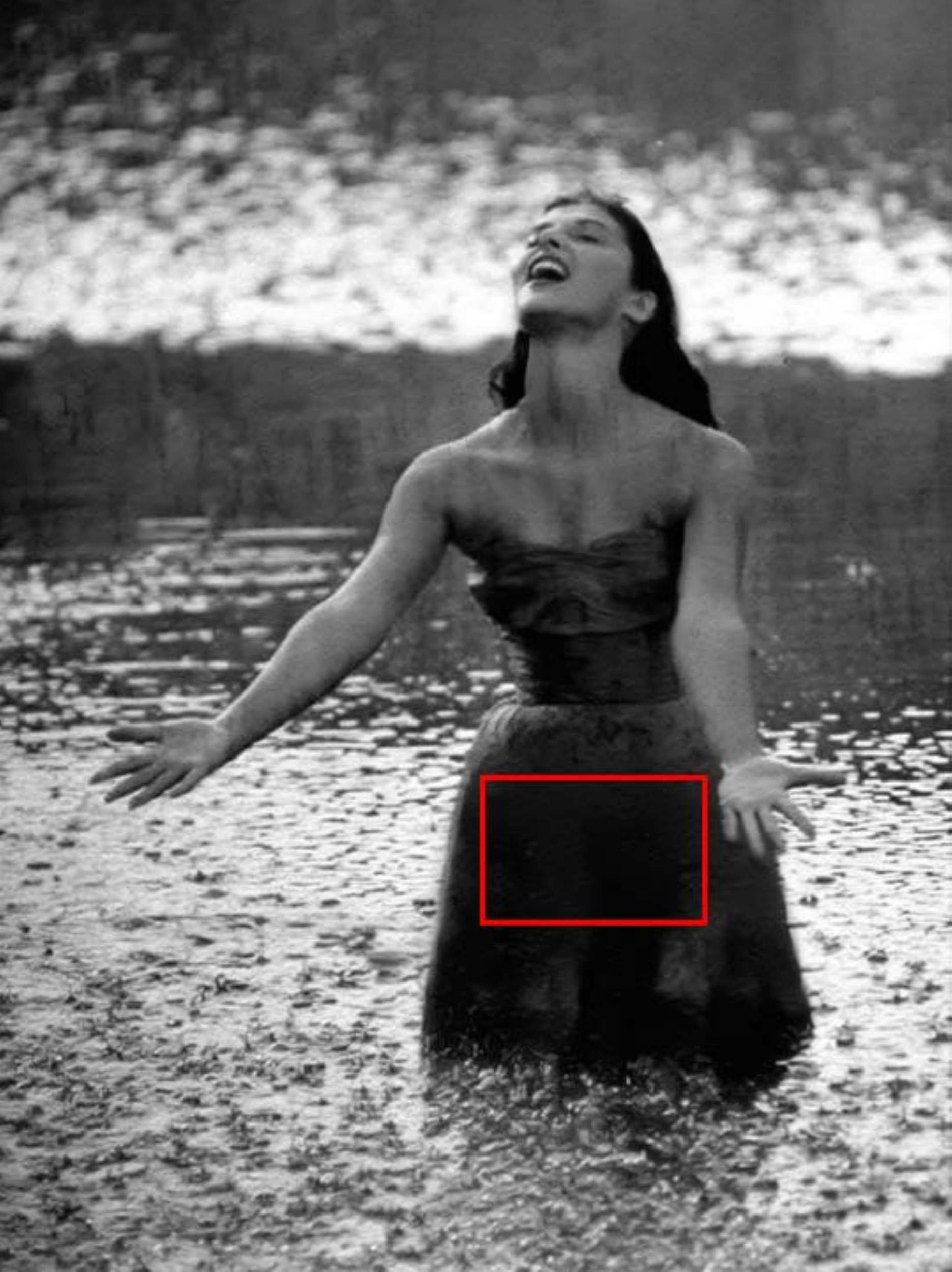}
\\

(a) Input &\hspace{-4mm} (b) DDN &\hspace{-4mm} (c) JORDER &\hspace{-4mm} (d) RESCAN &\hspace{-4mm} (e) PReNet &\hspace{-4mm} (f)~\textbf{Ours}
\\
\end{tabular}
\end{center}
\vspace{-4mm}
\caption{Results on two real-world examples of our method compared with four deep learning-based deraining methods.
Our results shown in (f) perform better than others visually.
Especially, for the first example, our method generates the clearest and cleanest result, while the other methods remain some obvious artifacts or rain streaks.
This demonstrates that our method performs superiorly to the state-of-the-art methods. \textbf{Please zoom in for the best visual comparisons!}}
\label{fig: real-world}
\vspace{-5mm}
\end{figure*}
\subsection{Results on real-world images}
\label{sec:Results on real-world images}
To evaluate the robustness of our method on real-world images, we also provide two examples on real-world rainy datasets in Fig.~\ref{fig: real-world}.
For the first example, our method generates the clearest and cleanest result, while the other methods remain some obvious artifacts or rain streaks.
For the second example, the other methods get unpleasing artifacts in the masked box, while our approach generates better clear results.
We provide more examples on both synthetic and real-world datasets in our supplemental materials.

%
\subsection{Ablation study}
\label{sec: Ablation study}

\begin{figure}[!h]
\begin{center}
\begin{tabular}{cccc}
\includegraphics[width = 0.23\linewidth]{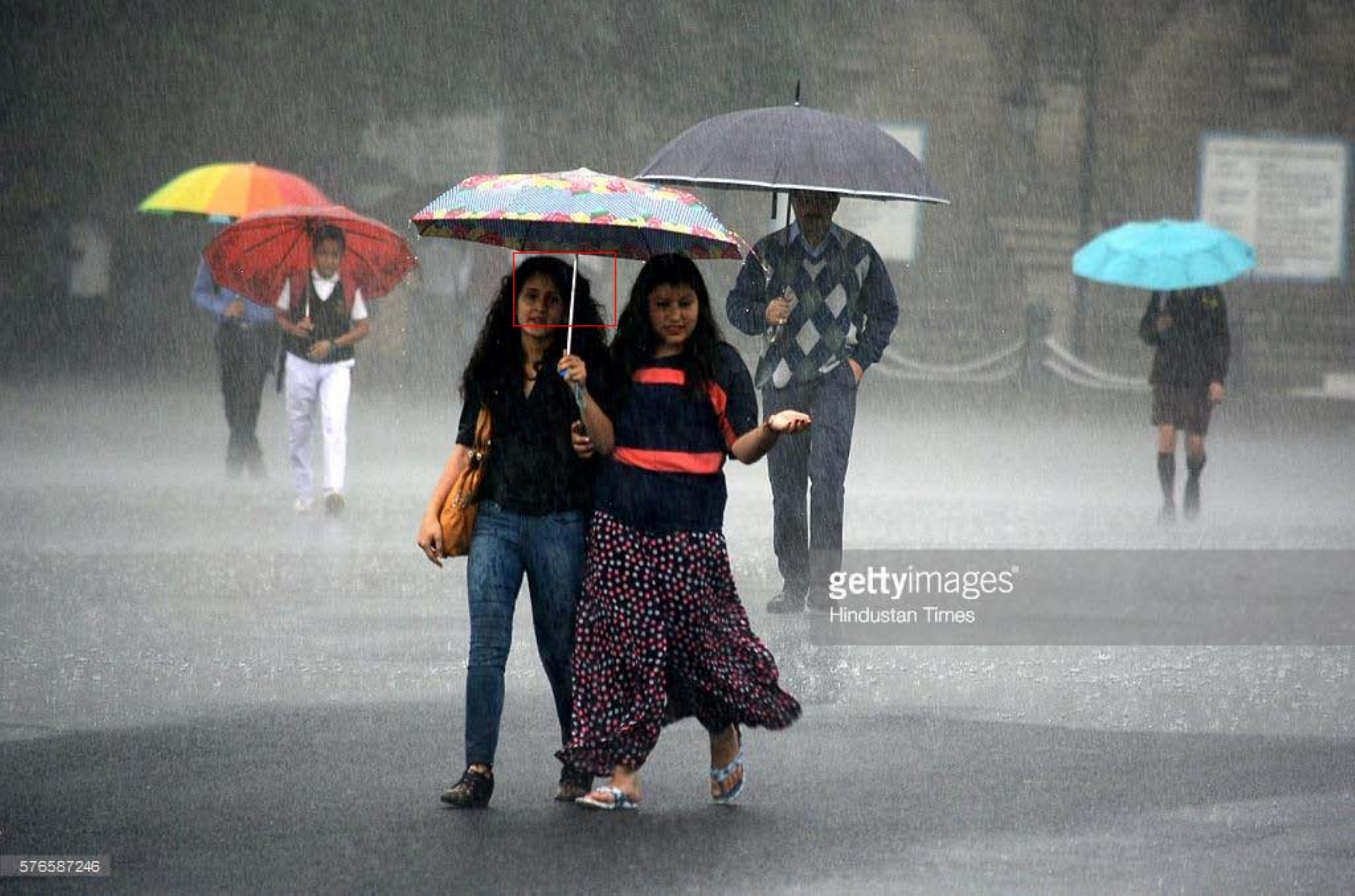} &\hspace{-3.5mm}
\includegraphics[width = 0.23\linewidth]{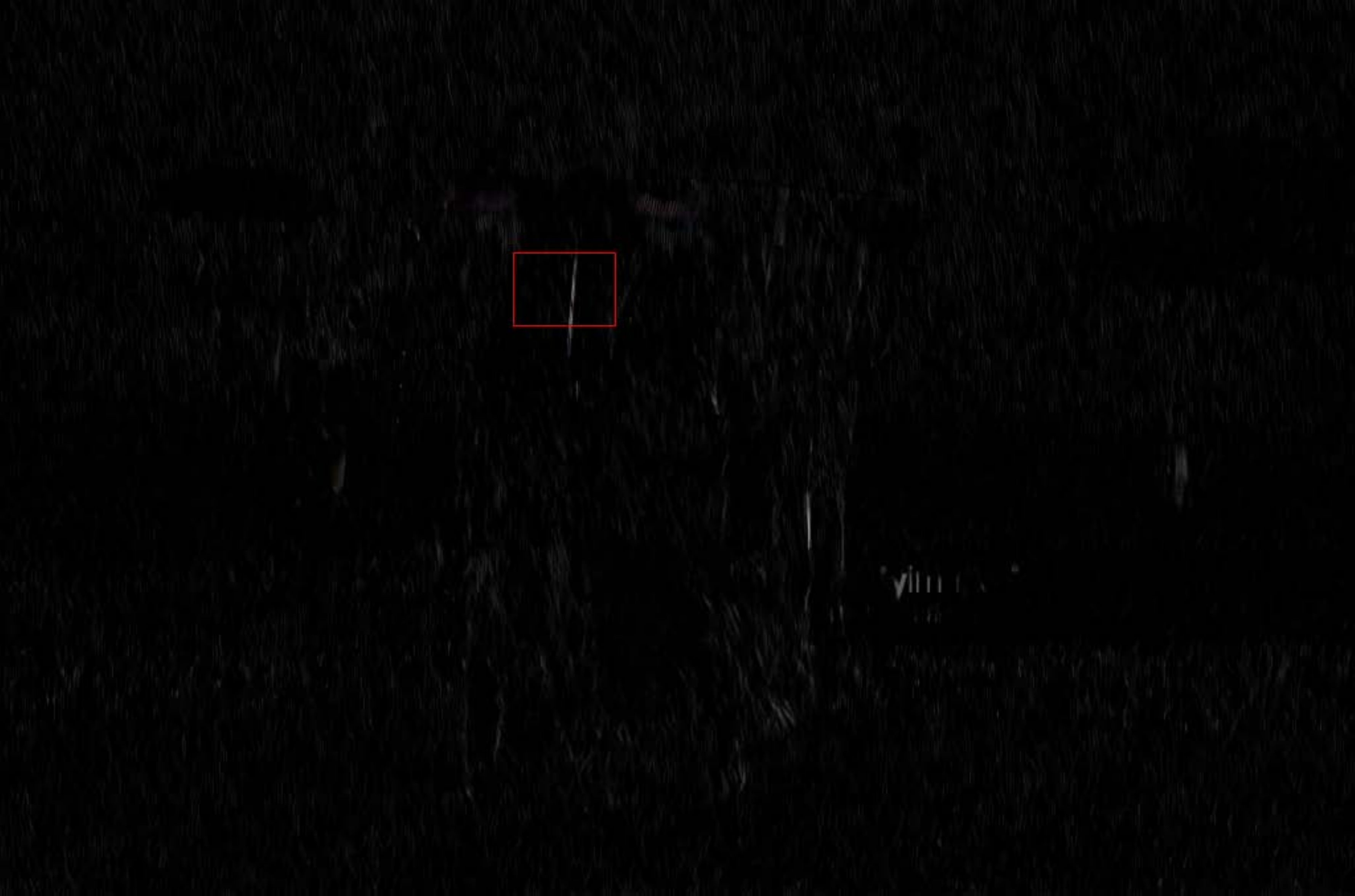} &\hspace{-3.5mm}
\includegraphics[width = 0.23\linewidth]{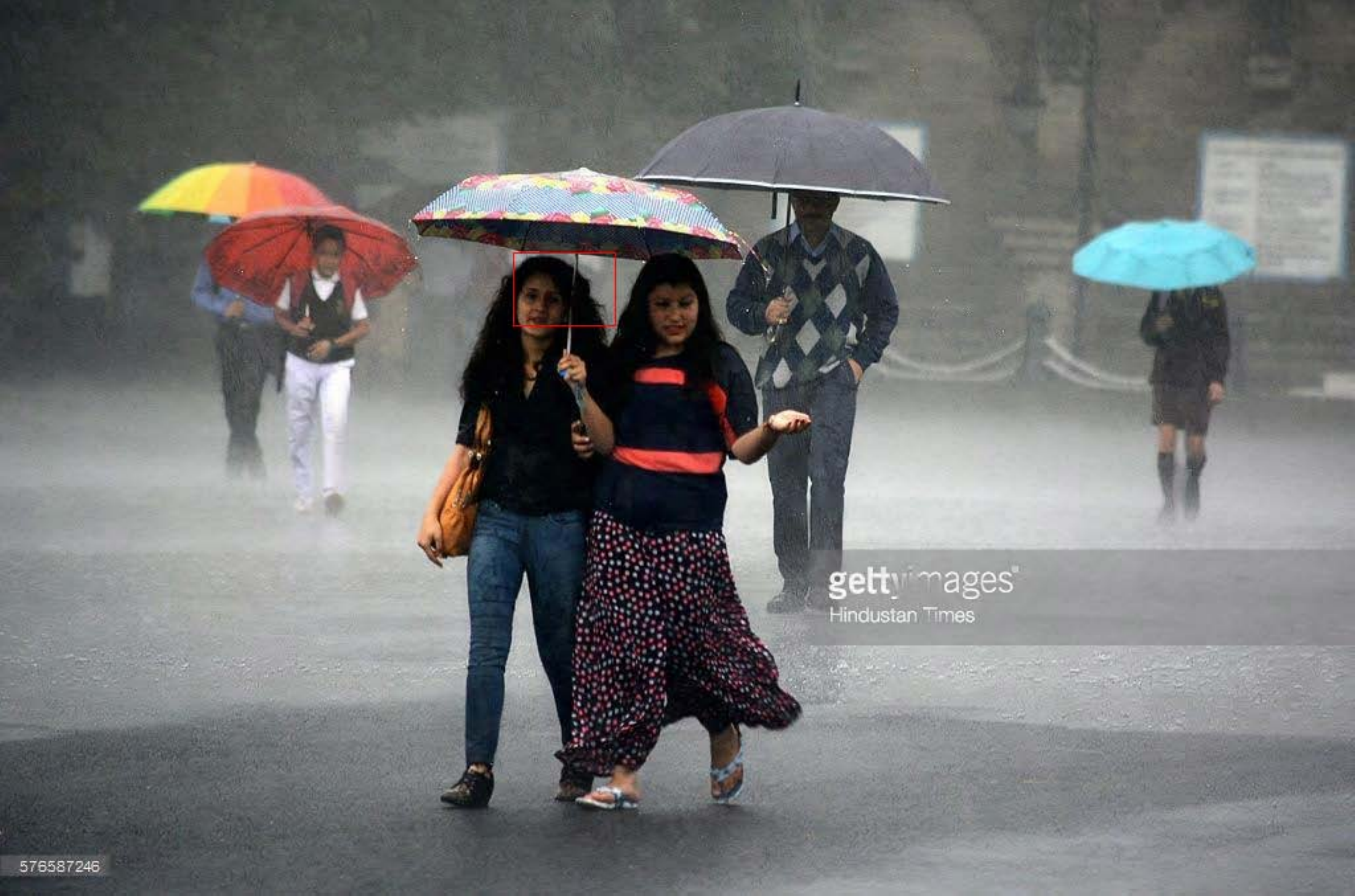} &\hspace{-3.5mm}
\includegraphics[width = 0.23\linewidth]{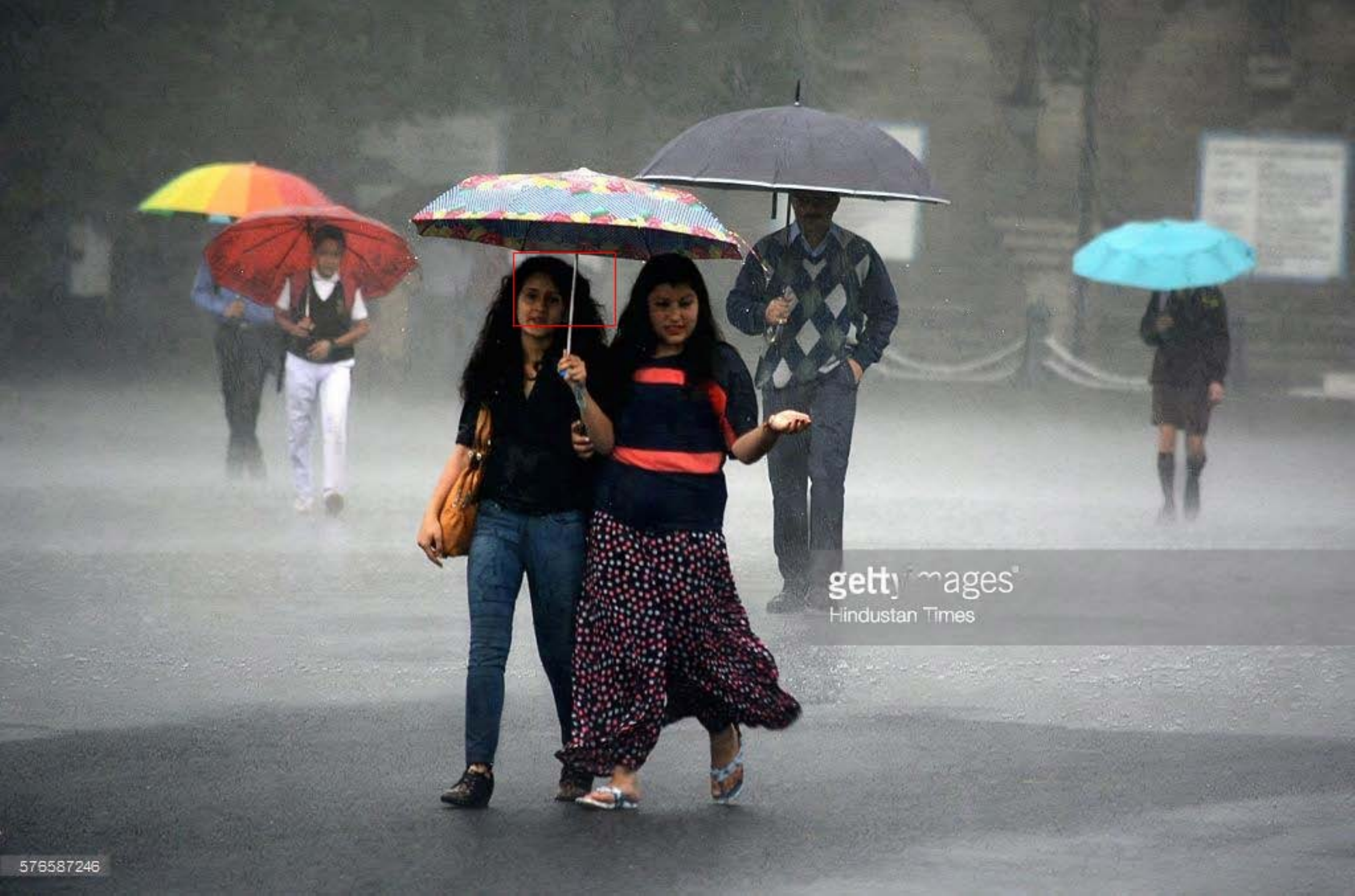}
\\
\includegraphics[width = 0.23\linewidth]{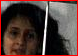} &\hspace{-3.5mm}
\includegraphics[width = 0.23\linewidth]{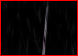} &\hspace{-3.5mm}
\includegraphics[width = 0.23\linewidth]{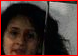} &\hspace{-3.5mm}
\includegraphics[width = 0.23\linewidth]{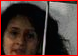}
\\
\includegraphics[width = 0.23\linewidth]{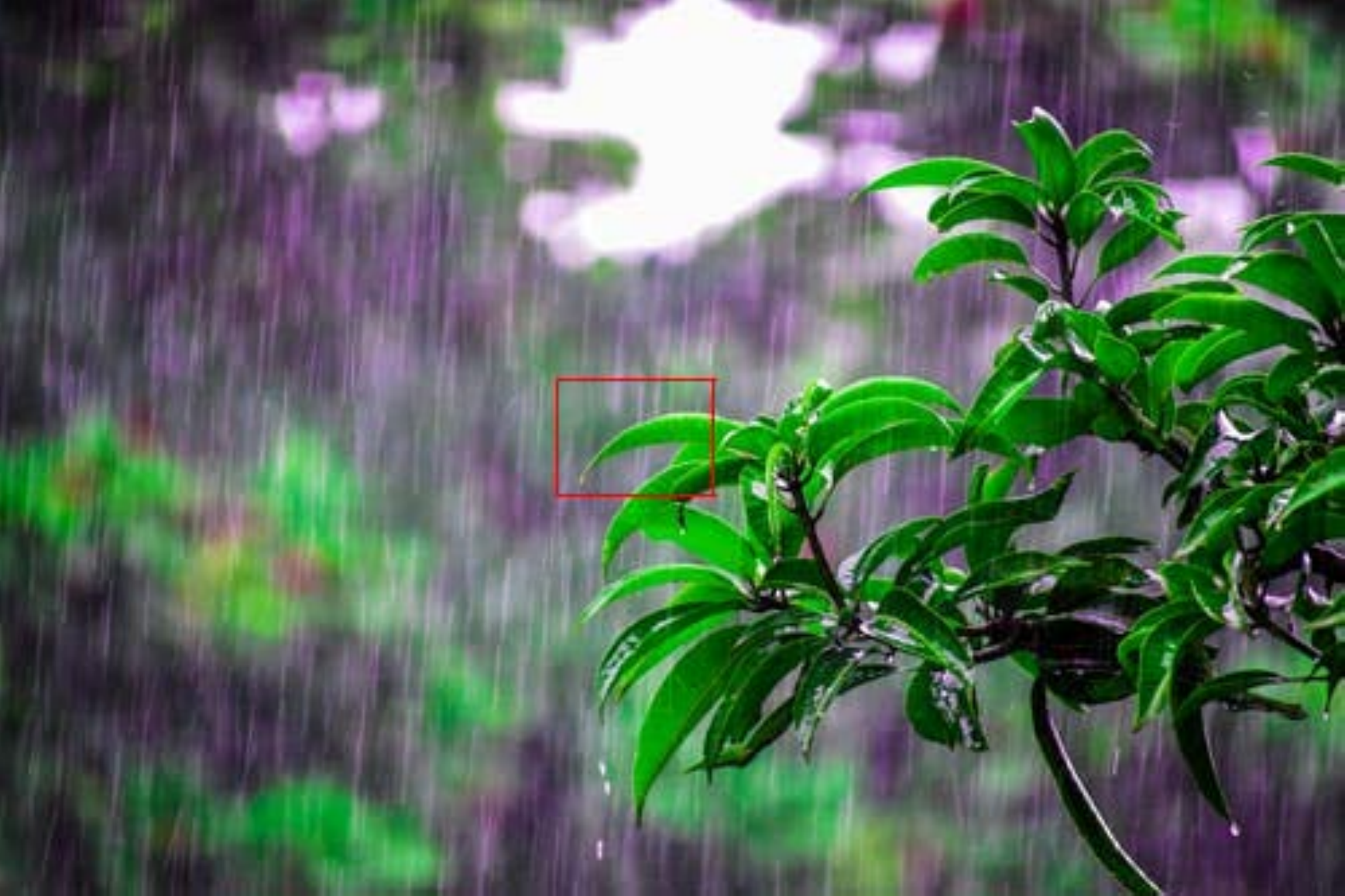} &\hspace{-3.5mm}
\includegraphics[width = 0.23\linewidth]{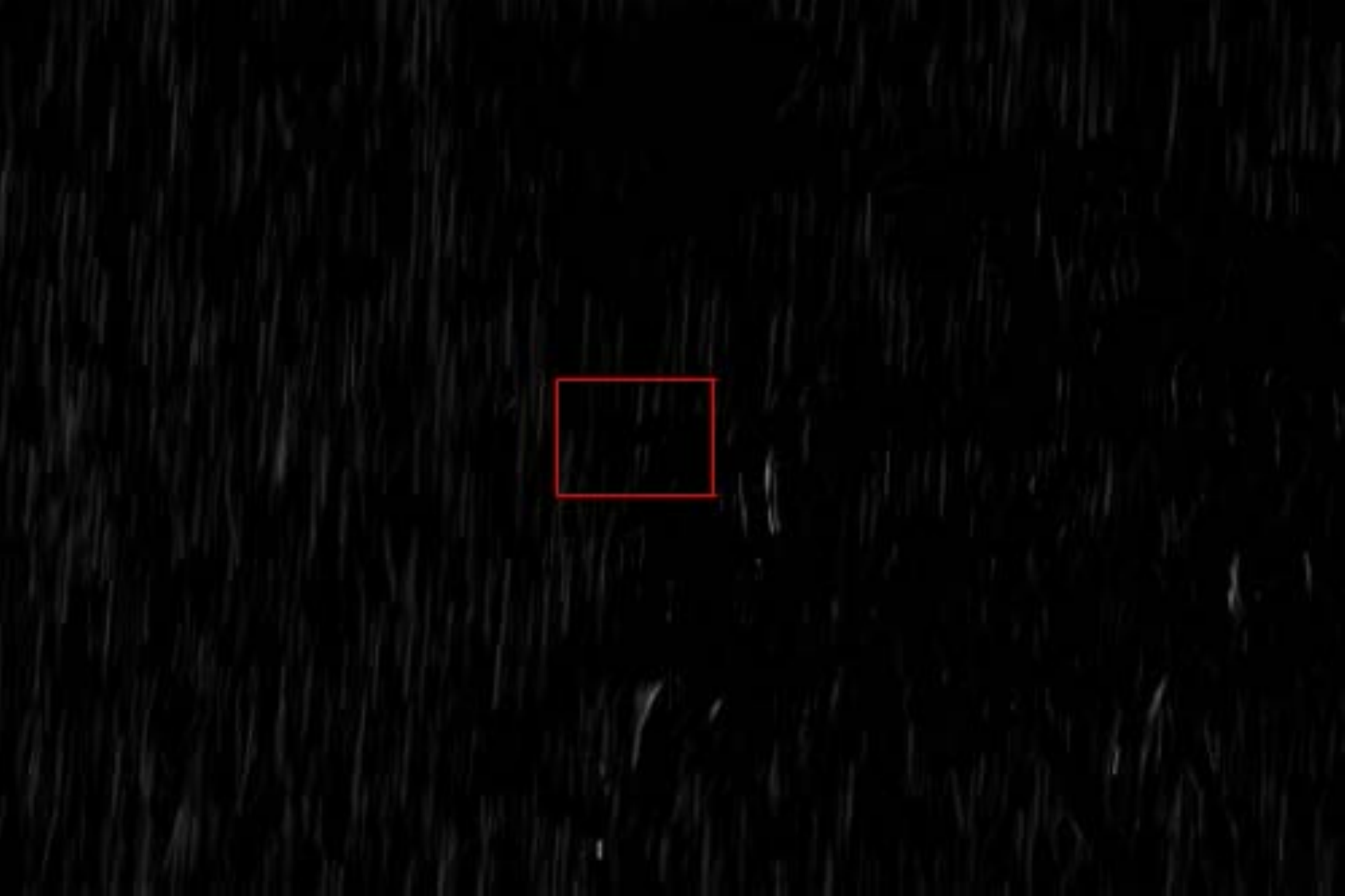} &\hspace{-3.5mm}
\includegraphics[width = 0.23\linewidth]{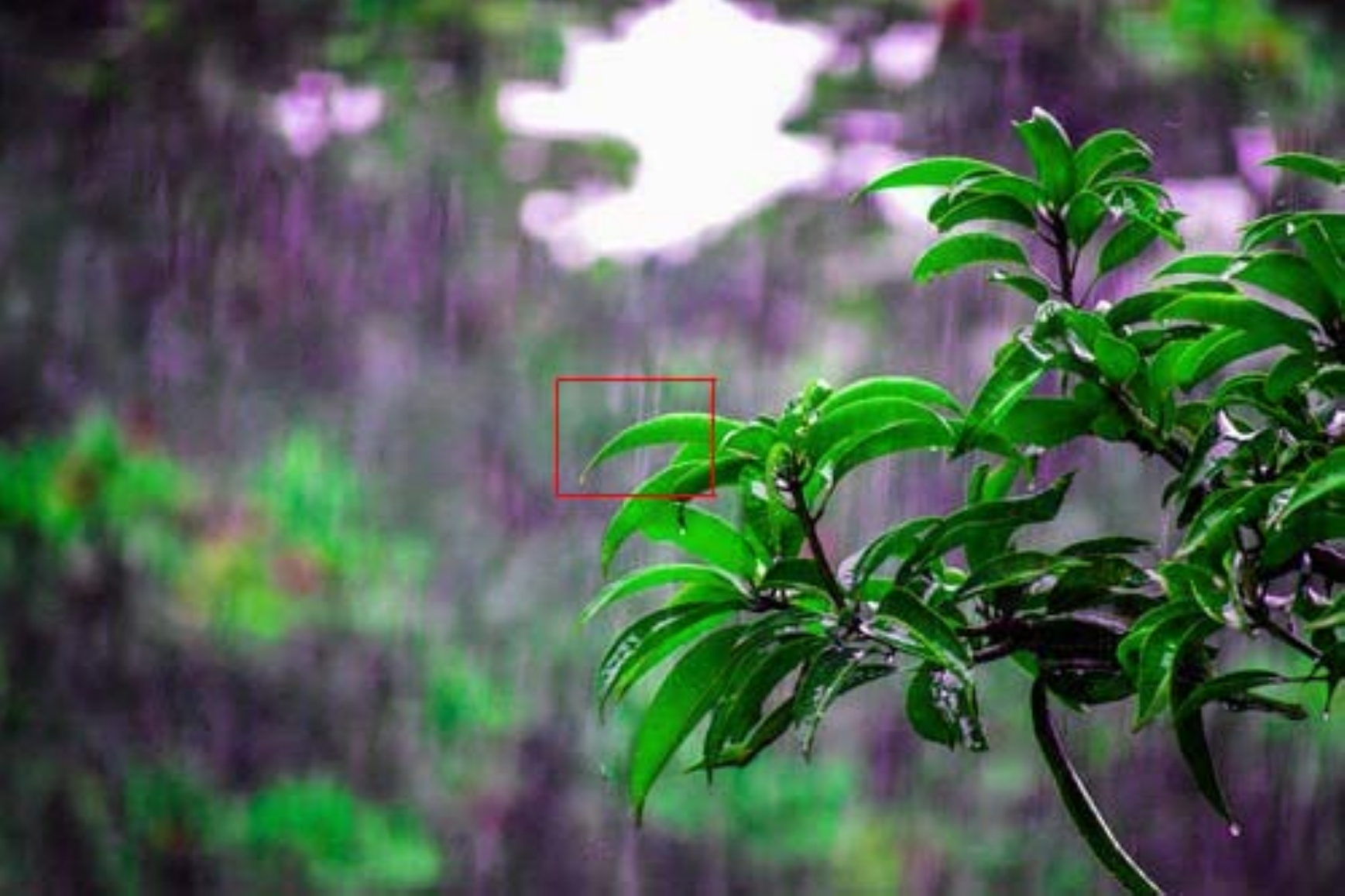} &\hspace{-3.5mm}
\includegraphics[width = 0.23\linewidth]{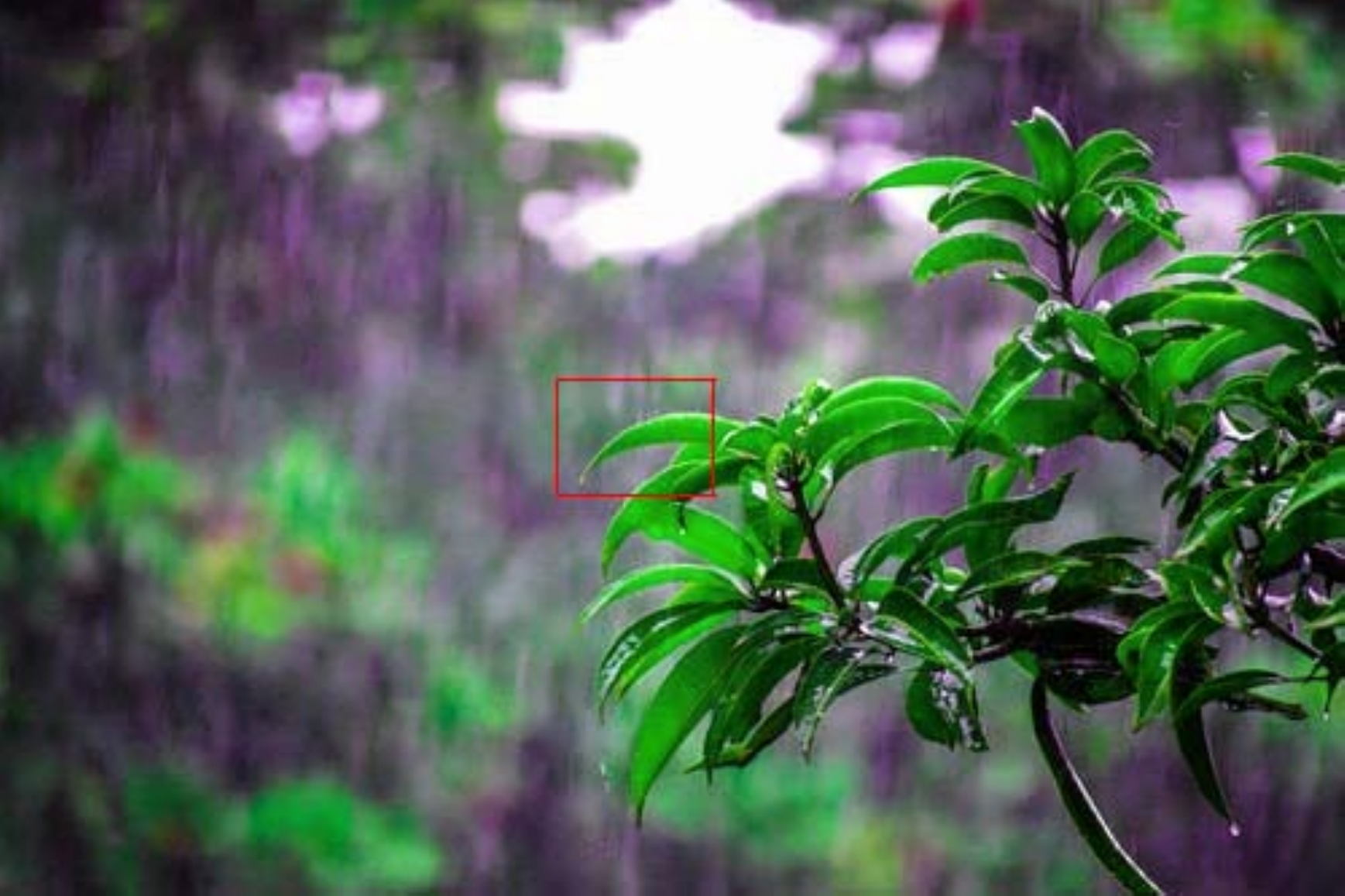}
\\
\includegraphics[width = 0.23\linewidth]{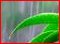} &\hspace{-3.5mm}
\includegraphics[width = 0.23\linewidth]{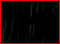} &\hspace{-3.5mm}
\includegraphics[width = 0.23\linewidth]{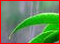} &\hspace{-3.5mm}
\includegraphics[width = 0.23\linewidth]{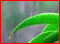}
\\
Input &\hspace{-4mm} Rain Streaks &\hspace{-4mm}  Rain-free   &\hspace{-4mm}  Output
\end{tabular}
\end{center}
\vspace{-5mm}
\caption{Outputs of different sub-networks on two real-world images.
Image in (b) is the estimated rain streaks, (c) and (d) show the outputs of rain-free network and guide-learning network, respectively.
}
\label{fig: sub-network}
\end{figure}

Exploring the effectiveness of multi-scale manner and the restraint of physical model in our network is meaningful.
So we design some experiments with different combinations of the proposed network components, such as three sub-networks, multi-scale structure, multi-stream dilation convolution, and $L_{p}$.
Tab.~\ref{tab:ablation pyramid} shows the comparative results and W and W/O mean whether using the multi-scale structure or not.
We can observe that the multi-scale manner boosts deraining performance on all models.
This illustrates that our designed multi-scale manner is meaningful.
Furthermore, Tab.~\ref{tab:ablation lp} compares the effectiveness of multi-stream dilation convolution and physical model constraint $L_{p}$.
Fig.~\ref{fig: sub-network} provides the outputs of three sub-networks on two real-world rainy images.
As we can see, the cropped patches in Fig.~\ref{fig: sub-network}~(d) perform better than Fig.~\ref{fig: sub-network}~(c) in detail and texture information, which demonstrates that the guide-learning network is effective in our proposed network.
\begin{table}[!h]
\vspace{-3mm}
\begin{center}
\caption{Ablation study on different models. The best results are marked in bold.}
\begin{tabular}{cccccc}
\hline
                     & Metric & M\_1       & M\_2        & M\_3         & \textbf{M\_4}   \\
\hline
\multirow{2}{*}{W}   & PSNR   & 28.63      & 28.56       & 28.62        & \textbf{28.97}  \\
\cline{2-6}
                     & SSIM   & 0.8949     & 0.8946      & 0.8968       & \textbf{0.9015} \\
\hline
\multirow{2}{*}{W/O} & PSNR   & 28.24      & 28.24       & 28.46        & \textbf{28.72}  \\
\cline{2-6}
                     & SSIM   & 0.8898     & 0.8900      & 0.8941       & \textbf{0.8986} \\
\hline
\end{tabular}
\label{tab:ablation pyramid}
\end{center}
\end{table}
%
\begin{itemize}
\vspace{-3mm}
\item M\_1: Only rain streaks network.
\item M\_2: Only rain-free network.
\item M\_3: Only input the estimated rain-free image to guide-learning network.
\item M\_4: (Default) the input is the concatenation of the estimated rain streaks and rain-free image to the guide-learning network.
\item R\_1: Our proposed network without multi-stream dilation convolution.
\item R\_2: Our proposed network without $L_{p}$.
\item R\_3: Our proposed network with multi-stream dilation convolution and $L_{p}$, i.e. our proposed final network.
\end{itemize}
$L_{p}$ is the $L_1$-norm of subtraction between the rainy image and the direct sum of the estimated two images from the first two sub-networks.
%

\begin{table}[!h]
\centering
\caption{Analysis on the effectiveness of multi-stream dilation convolution and physical model constraint. The best results are marked in bold.}
\scalebox{0.99}
{
\begin{tabular}{cccc}
\hline
Metric     & R\_1        & R\_2        &\textbf{R\_3}     \\
\hline
PSNR       & 28.95       & 28.92       &\textbf{28.97}    \\
\hline
SSIM       & 0.9011      & 0.9012      &\textbf{0.9015}   \\
\hline
\end{tabular}
}
\label{tab:ablation lp}
\end{table}
\vspace{-5mm}
\section{Conclusion}
In this paper, we propose an effective method to handle single image deraining.
Our network is based on the rainy physical model with guide-learning and the experiments demonstrate the physical model constraint and guide-learning are meaningful.
Multi-Scale Residual Block is proposed and verified to boost the deraining performance.
Quantitative and qualitative experimental results on both synthetic datasets and real-world datasets demonstrate the favorable of our network for single image deraining.
\section*{Acknowledgement}
This work was supported by National Natural Science Foundation of China [grant numbers 61976041]; National Key R\&D Program of China [grant numbers 2018AAA0100301]; National Science and Technology Major Project [grant numbers 2018ZX04041001-007, 2018ZX04016001-011].
%
{
\small
\bibliographystyle{IEEEbib}
\bibliography{icme20-reference}
}
\end{document}